\documentclass[aps,nofootinbib,amsmath]{revtex4}

\usepackage{amsfonts}
\usepackage{amsmath}
\usepackage{amssymb}
\usepackage{graphicx}
\usepackage{color}

\usepackage{multirow} 
\usepackage{tabularx}

\def\bibsection{\section*{\refname}} 

\begin{document}
\title{Neutrino mass matrix in neutrino-related processes}

\author{M. I. Krivoruchenko}
\email{mikhail.krivoruchenko@itep.ru}
\affiliation{National Research Centre ''Kurchatov Institute'', Moscow, Russia}
\author{F. \v Simkovic}
\email{fedor.simkovic@fmph.uniba.sk}
\affiliation{Faculty of Mathematics, Physics and Informatics$\mathrm{,}$ Comenius University in Bratislava, Bratislava, Slovakia \\
Institute of Experimental and Applied Physics, Czech Technical University in Prague, Prague, Czech Republic}

\begin{abstract}
Techniques are developed for constructing amplitudes of neutrino-related processes in terms of neutrino mass matrix, 
with no reference to neutrino mixing matrix.
The amplitudes of neutrino oscillations in vacuum and medium, quasi-elastic neutrino scattering, electron capture, $\beta$ decays and double-$\beta$ decays are considered.
The proposed approach makes extensive use of Frobenius covariants within the framework of Sylvester's theorem on matrix functions. 
The in-medium dispersion laws in baryon matter are found in terms of elementary functions for three flavors of Majorana neutrinos  
as an application of the developed formalism. 
The in-medium dispersion laws for Dirac neutrinos can be determined in the general case by searching for roots 
of a polynomial of degree 6.
In the rest frame of baryon matter, the minimum energy of both Majorana and Dirac neutrinos is achieved at a finite neutrino momentum proportional to the mean-field potential. 
In such cases, Dirac neutrinos occupy a hollow Fermi sphere at zero temperature and low densities.
Fitting experimental data in terms of neutrino mass matrix can provide better statistical accuracy in determining neutrino mass matrix compared to methods using neutrino mixing matrix at intermediate stages.
\end{abstract}
\maketitle

\section{Introduction}
\renewcommand{\theequation}{I.\arabic{equation}}
\setcounter{equation}{0}

The predictions of the Standard Model (SM) are effectively confirmed by a vast number of experiments. 
The first discrepancy with SM was discovered in 1998 as a result of observations of atmospheric neutrino 
oscillations \cite{Fukuda:1988}, proving the existence of finite neutrino masses. By adding the right-handed neutrinos to SM while maintaining the renormalizability of the model, it is possible to generate Dirac neutrino masses. When the non-renormalizable $d=5$ interaction of left-handed leptons with the Higgs boson is turned on, masses of Majorana neutrinos appear \cite{Weinberg:1979}. A plausible hierarchy of masses is suggested by the seesaw mechanism involving both Dirac and Majorana neutrinos \cite{Bilenky:2018}. 
The mechanism of neutrino mass generation by the spontaneous chiral symmetry breaking is also discussed \cite {Babic:2021}. Only in isolated cases 
the theory encounters challenges in adjacent SM sectors, 
the interpretation of measurements of the muon's anomalous magnetic moment \cite{Labe:2022} 
or the decays of beauty mesons \cite{LHCb:2022}.

In the searches for physics beyond SM, neutrinos occupy a unique place.
The mass matrices of Majorana and Dirac neutrinos have 9 and 7 free parameters, respectively, already in the simplest case of three generations \cite{Kobzarev:1982,Ismael:2021}. 
This is roughly half of the total number of SM free parameters.
The elements of neutrino mass matrix may contain a great deal of information on potential SM extensions.

Cross-sections of processes involving neutrinos, probabilities of oscillations and decay rates
are usually expressed in terms of the Pontecorvo--Maki--Nakagawa--Sakata (PMNS) mixing matrix and 
the diagonal neutrino masses. 
Significant progress in the experimental determination of the elements of the mixing matrix and obtaining restrictions on neutrino masses has been achieved over the past decade due to the analysis of neutrino oscillations, including data from atmospheric experiments, accelerator data on the modes of appearance and disappearance of neutrinos, and data from reactor experiments \cite{deSalas:2017kay,Capozzi:2017ipn}. 
The elements of the mass matrix, in principle, are accessible to experimental determination in terms of the mixing matrix and the diagonal neutrino masses, 
but contain uncertainties associated with arbitrariness of the absolute mass scale. 
An indirect approach to the determination of the mass matrix can lead to loss of information at intermediate stages of statistical analysis. 
In this paper, we show that the processes associated with neutrinos are explicitly described in terms of the neutrino mass matrix, without using a mixing matrix.
The existence of such representations makes it possible to fit the elements of neutrino mass matrix directly from experimental data, which should provide an advantage when conducting statistical analysis
\textcolor{black}{and, potentially, distinguish the Majorana mass matrix from the Dirac one.}

Another purpose of this paper is to point out effective techniques of matrix algebra that significantly simplify the work with neutrinos of three or more flavors. 
The paper uses extensively Sylvester's theorem on the representation of matrix functions \cite{Roger:1991}, 
Frobenius covariants, which determine projection operators in terms of the matrix under consideration, 
and trace identities \cite{Kondratyuk:1992,Brown:1994}, 
which provide explicit solutions to the Faddeev–LeVerrier algorithm \cite{Frame:1949,Faddeev:1972}. 
The described techniques allow to effectively construct the neutrino dispersion laws. The cases of $n \leq 4$ flavors of Majorana neutrinos in a baryon matter and $n \leq 4$ flavors of Dirac neutrinos in a baryon matter of arbitrary composition are solved in terms of elementary functions.

The outline of the paper is as follows. 
Neutrino oscillations in a vacuum are discussed in the next section. 
Neutrino oscillations in the medium are covered in Section III. Majorana neutrinos and Dirac neutrinos are considered separately. 
The neutrino dispersion laws are derived, and it is demonstrated that the mass matrix explicitly describes the physical amplitudes of neutrino propagation in all cases considered.
The neutrino dispersion law in the medium takes an unusual form: the minimum neutrino energy is achieved at a finite momentum.
Other processes covered briefly in Section IV include tritium $\beta$ decay, electron capture, double-$\beta$ decays, and quasi-elastic neutrino scattering. In all circumstances, probabilities are explicitly expressed in terms of the neutrino mass matrix rather than the mixing matrix.


\section{Neutrino oscillations in vacuum}
\renewcommand{\theequation}{II.\arabic{equation}}
\setcounter{equation}{0}

\subsection{Majorana neutrino}

Let us first consider the problem of oscillations of Majorana neutrinos. The Lagrangian of free Majorana neutrinos is written as
\begin{equation} \label{LM}
\mathcal{L}_{M}=\frac{1}{2}\bar{\nu}_{\alpha} i\hat{\nabla}\nu _{\alpha}-\frac{1}{2}%
\bar{\nu}_{\alpha} \left( M_{\alpha \beta}+i\gamma _{5}N_{\alpha \beta}\right) \nu _{\beta}.
\end{equation}
Here  $\nu _{\alpha}$ are the real four-component spinors of Majorana neutrinos, the gamma matrices are defined in the Majorana representation
where $\gamma^*_{\mu} = - \gamma_{\mu}$, $\gamma^T_{\mu} = - \gamma^{\mu}$, 
$\gamma^*_{5} = \gamma^T_{5} = -\gamma_{5}$, 
$\gamma^{\dagger}_{5} = \gamma_{5}$ and
$\nu _{\alpha}^c = \nu _{\alpha}^* = \nu _{\alpha}$, $\bar{\nu}_{\alpha}^c = {\nu}_{\alpha}^{T}\gamma_{0}$.
The left- and right-handed neutrinos are defined by $\nu _{\alpha L} = \Pi_- \nu _{\alpha}$
and $\nu _{\alpha  R} = \Pi_+ \nu _{\alpha}$, respectively, where $\Pi_\pm = (1 \pm \gamma_5)/{2}$ are the projection operators onto states with the definite chirality.
The transformation properties of the spinor bilinear forms determine the
mass term symmetry. We denote by $\Gamma_A$ the matrices $ 1,i\gamma _{5},\gamma ^{\mu },\gamma _{5}\gamma ^{\mu},\sigma ^{\mu \nu }$, defined in the same way as in ~\cite{Bjorken:1964}.
Taking into account $\gamma_0 \Gamma_A^\dagger \gamma_0 = \Gamma_A$ and the fermionic character of the fields, for both Dirac and Majorana neutrinos
\begin{equation}
\bar{\nu}_{\alpha}\Gamma \nu _{\beta}=-\bar{\nu}_{\beta}^{c}\gamma ^{0}(\gamma
^{0}\Gamma_A )^{T}\nu _{\alpha }^{c}=\eta \bar{\nu}_{\beta}^{c}\Gamma_A \nu _{\alpha}^{c}.
\end{equation}
The sign $\eta $ is read from the equation (cf. Appendix B in \cite{Krivoruchenko:2011})%
\begin{equation}
-\gamma ^{0}(\gamma ^{0}\left(
\begin{array}{l}
1 \\
i\gamma _{5} \\
\gamma ^{\mu } \\
\gamma _{5}\gamma ^{\mu } \\
\sigma ^{\mu \nu }%
\end{array}%
\right) )^{T}=\left(
\begin{array}{l}
1 \\
i\gamma _{5} \\
-\gamma ^{\mu } \\
\gamma _{5}\gamma ^{\mu } \\
-\sigma ^{\mu \nu }%
\end{array}%
\right) .  \label{xsxsx}
\end{equation}
The conditions $\bar{\nu}_{\alpha }\nu _{\beta}=\bar{\nu}_{\beta}\nu _{\alpha }$
and $\bar{\nu}_{\alpha}i\gamma _{5}\nu _{\beta}=\bar{\nu}_{\beta}i\gamma _{5}\nu_{\alpha }$
valid for Majorana fields imply that the mass matrices $M$ and $N$ are symmetric.
The hermiticity of the Lagrangian ensures the hermiticity of $\mathbb{M}$ and $\mathbb{N}$. Since they are symmetric, 
they are real.
We define for $n$ flavors a $4n\times 4n$ real matrix
\begin{equation} \label{matrix12}
\hat{\mathbb{M}} = \mathbb{M} +i\gamma _{5}\mathbb{N}
\end{equation}
and  $n\times n$  complex symmetric matrices
\begin{equation} \label{matrix}
\mathbb{M}_{\pm }= \mathbb{M} \pm i\mathbb{N}.
\end{equation}
One can verify that
$\hat{\mathbb{M}}^\dagger =  \hat{\mathbb{M}}^T = \gamma_0 \hat{\mathbb{M}} \gamma_0= \mathbb{M}-i\gamma _{5}\mathbb{N}$
and $\mathbb{M}_{+}^\dagger =  \mathbb{M}^*_{+} = \mathbb{M}_{-}$. 
We also use notations $\mathbb{M}_{R} = \mathbb{M}_{+}$, $\mathbb{M}_{L} = \mathbb{M}_{-}$
and write $\mathbb{M}_{\pm 1} $ for $ \mathbb{M}_{\pm}$ and $\Pi_{\pm 1}$ for $\Pi_{\pm}$.
The matrices $\mathbb{M}$ and $\mathbb{N}$ are specified by $2\times (n(n+1)/2)$ real parameters. 
Phase rotations of Majorana spinors $\nu_{\alpha} \to \exp(i\gamma_5 \phi_{\alpha}) \nu_{\alpha}$ 
keep the kinetic term in the Lagrangian and the weak neutral current $j_{\mu}^{\mathrm{NC}}$ unchanged but modify the mass matrix  and introduce the phase multipliers into the weak charged current $j_{\mu}^{\mathrm{CC}}$. 
The phase multipliers in $j_{\mu}^{\mathrm{CC}}$ are absorbed further by the charged leptons. The number of remaining free parameters in the mass matrix equals $n^2$. 

In the representation of two-component Weyl spinors $\nu _{\alpha L}$, 
the Lagrangian (\ref{LM}) has the form
\begin{equation} \label{SM}
\mathcal{L}_{M}=\frac{1}{2}\bar{\nu}_{\alpha L}i\hat{\nabla}\nu_{\alpha L}-\frac{1}{2}
\bar{\nu}_{\alpha L}M_{\alpha \beta}^{M}\nu_{\beta L}^{c}+ \mathrm{h.c.},
\end{equation}%
where  $\nu _{L \alpha}^c =  (\nu _{L \alpha})^c$ and 
$M^{M}_{\alpha \beta} =  M_{\alpha \beta} + i N_{\alpha \beta}$ 
is a symmetric complex $n\times n$ matrix  $\mathbb{M}_{R}$ defined above. 
It is useful to identify phases of the mass matrix elements:
$M^{M}_{\alpha \beta} = | M^{M}_{\alpha \beta} | e^{i\phi^{\prime}_{\alpha \beta}}.$
The chiral rotations $\nu_{\alpha} \to \exp(- i\gamma_5 \phi^{\prime}_{\alpha \alpha}/2) \nu_{\alpha}$ enable the diagonal elements of $\mathbb{N}$ to be set to zero and the diagonal elements of  $\mathbb{M}_R$ to be considered as real 
and positive. For three generations
\begin{equation} \label{massmatrix} 
\mathbb{M}_{R}=\left\Vert 
\begin{array}{lll}
M^{M}_{ee} & |M^{M}_{e\mu } |e^{i\phi_{e\mu }} & |M^{M}_{e\tau } |e^{i\phi_{e\tau }} \\ 
|M^{M}_{e\mu } |e^{i\phi_{e\mu }} & M^{M}_{\mu \mu } & |M^{M}_{\mu \tau } |e^{i\phi_{\mu
\tau }} \\ 
|M^{M}_{e\tau }|e^{i\phi_{e\tau }} & |M^{M}_{\mu \tau }|e^{i\phi_{\mu \tau }} & 
M^{M}_{\tau \tau }%
\end{array}
\right\Vert, 
\end{equation}
where $\phi_{\alpha \beta} = \phi^{\prime}_{\alpha \beta} - (\phi^{\prime}_{\alpha \alpha} + \phi^{\prime}_{\beta \beta})/2$.
Ref.~\cite{Pal:2011} discusses in detail the relationship between Weyl spinors and Majorana and Dirac bispinors.

The free Lagrangian (\ref{LM}) is invariant with respect to the charge conjugation: 
$\nu _{\alpha }\rightarrow \nu _{\alpha }^{c}=\nu _{\alpha }^{\dagger}$, 
however, due to the special type of the mass matrix, it is non-invariant
with respect to $P$-parity transformations
$\nu _{\alpha } \rightarrow  \mathcal{P}\nu _{\alpha }\mathcal{P}^{-1}=i\gamma _{0}\nu_{\alpha }$, 
and time reversal
$\nu _{\alpha }  \rightarrow  \mathcal{T}\nu _{\alpha }\mathcal{T}^{-1}=\gamma _{0}\gamma
_{5}\nu _{\alpha }$,
When weak interactions are turned on, only 
the combined CPT 
symmetry survives.
$CP$ invariance would call for $\mathbb{N} = 0$ and forbid phase rotations with the $\gamma_5$ matrix, which would leave $n(n+1)/2$ free parameters in the mass matrix.

The Feynman propagator has the form
\begin{equation} \label{SFcoord}
\hat{\mathbb{S}}_F(x-y) = -i\langle 0|T{\nu}(x)\otimes\bar{\nu}(y)|0 \rangle = \int \frac{d^4p}{(2\pi)^4}e^{-ip(x-y)}\hat{\mathbb{S}}_F(p),
\end{equation}
where
\begin{eqnarray}
\hat{\mathbb{S}}_F(p) = \frac{1}{\hat{p} - \hat{\mathbb{M}}} \label{SF}
\end{eqnarray}
is determined by the quadratic form of the Lagrangian. 
The matrix $\hat{p} - \hat{\mathbb{M}}$ has the dimension $4n$. The energy eigenvalues are grouped by pairs $\pm \sqrt{m_i^2 + \mathbf{p}^2}$ for different masses $m_i$ ($i = 1,..,n$). The additional doubling is due to the spin degeneracy. 
Using the identity $A^{-1}=B(AB)^{-1}=(BA)^{-1}B$
for $A = \hat{p} - \hat{\mathbb{M}}$ and $B = \hat{p} + \hat{\mathbb{M}}^\dagger$ and $\hat{p}\hat{\mathbb{M}} = \hat{\mathbb{M}}^\dagger\hat{p}$, we obtain
\begin{eqnarray} \label{SF2}
\hat{\mathbb{S}}_F(p) = (\hat{p} + \hat{\mathbb{M}}^\dagger)\frac{1}{p^{2}-\hat{\mathbb{M}}\hat{\mathbb{M}}^\dagger} 
= \frac{1}{p^{2}-\hat{\mathbb{M}}^\dagger \hat{\mathbb{M}}} (\hat{p} + \hat{\mathbb{M}}^\dagger).
\end{eqnarray}

The chirality-state projections can be simplified with the use of the relationships
$f(\hat{\mathbb{M}}          )\Pi_\varsigma = f(\mathbb{M}_{\varsigma}    )\Pi_\varsigma$ and 
$f(\hat{\mathbb{M}}^\dagger)\Pi_\varsigma         = f(\mathbb{M}_{- \varsigma}    )\Pi_\varsigma$, 
which hold true for any function $f(x)$ ($\varsigma = \pm 1$).
The projection of $\hat{\mathbb{S}}_F(p)$ involved in the creation of a neutrino with chirality $- \varsigma$ has the form:
\begin{eqnarray}
\hat{\mathbb{S}}_F(p) \Pi_{\varsigma} = (\hat{p} + \mathbb{M}_{- \varsigma}) \frac{1}{p^2 
- \mathbb{M}_{\varsigma}\mathbb{M}_{- \varsigma}}\Pi_{\varsigma}.
\end{eqnarray}
The advantage of this representation is that in the denominator there is a $n \times n$ matrix multiplied by
a unit matrix in the space of bispinors, while the original expression contains a $4n \times 4n$
matrix with the mixed bispinor and flavor indices.
The mass spectrum is determined by poles of the propagator, in this instance by zeros of the characteristic polynomial
\begin{eqnarray} \label{character}
p(\lambda) = \det||\lambda - \mathbb{M}_{\varsigma}\mathbb{M}_{- \varsigma}|| = 0.
\end{eqnarray}
For three flavors, $p(\lambda)$ is a polynomial of the third degree that has, respectively,
either three real roots, or one real and two complex roots.
The mass spectrum does not depend on $\varsigma$, because
$\det||\lambda - \mathbb{M}_{L} \mathbb{M}_{R}|| =
\det||\lambda - \mathbb{M}_{R}\mathbb{M}_{L}||$.
Given that $(\mathbb{M}_{L} \mathbb{M}_{R})^* = \mathbb{M}_{R}\mathbb{M}_{L}$,
the complex conjugation of $p(\lambda)$ gives the characteristic equation $p(\lambda^*) = 0$. Thus, either all three roots are real,
or one root is real, and two roots have imaginary parts of opposite sign.
The product $\mathbb{M}_{L} \mathbb{M}_{R}$, however, is a Hermitian matrix.
Hence all eigenvalues of $\mathbb{M}_{L} \mathbb{M}_{R}$ are real.
Let $\lambda_i = m_i^2$ be one of them. There exists a normalized vector $|i \varsigma \rangle$ such that
\begin{equation}  \label{eigen}
\mathbb{M}_{\varsigma}\mathbb{M}_{- \varsigma} | i \varsigma \rangle = \lambda_i |i \varsigma \rangle.
\end{equation}
Multiplying on the left by $\langle i \varsigma |$ yields
$\lambda_i = \langle i \varsigma | (\mathbb{M}_{- \varsigma})^\dagger \mathbb{M}_{- \varsigma} |i \varsigma \rangle > 0$.
Furthermore, if $|i \varsigma \rangle$ is a normalized eigenvector of $\mathbb{M}_{\varsigma}\mathbb{M}_{- \varsigma}$, then
\begin{equation} \label{bases}
|i - \varsigma \rangle = \frac{1}{m_i} \mathbb{M}_{- \varsigma} |i \varsigma \rangle
\end{equation}
is the normalized eigenvector of $\mathbb{M}_{- \varsigma} \mathbb{M}_{\varsigma}$ with the same eigenvalue $\lambda_i$.
In the basis of flavor states $|\alpha \rangle$, the coordinates of these eigenvectors appear to be complex conjugate: $\langle \alpha |i\varsigma \rangle ^* = \langle \alpha |i - \varsigma \rangle$.
One can check the validity of this relation by writing Eq.~(\ref{eigen}) in the flavor basis. After performing a complex conjugation, an equation follows for the eigenvectors of $\mathbb{M}_{- \varsigma} \mathbb{M}_{\varsigma}$, which means the eigenvectors of $\mathbb{M}_{\varsigma} \mathbb{M}_{- \varsigma}$ and $\mathbb{M}_{- \varsigma} \mathbb{M}_{\varsigma}$ differ 
by phase factors only. The phase factors of $|i \varsigma \rangle$, however, can always be chosen so that the relationship 
\begin{equation}  \label{symm}
\langle \alpha |i \varsigma \rangle ^* = \langle \alpha |i - \varsigma \rangle
\end{equation}
is fulfilled.
The mixing of Majorana neutrinos is described thereby in terms of a single mixing matrix. 
Equation (\ref{bases}) is equivalent to the condition $\langle i-|\mathbb{M}_{L}|j + \rangle = m_i \delta_{ij}$, which shows that for $n$ flavors the mass matrix $\mathbb{M}_{L}$ is diagonalized by two unitary matrices: 
\begin{equation} \label{VMU}
\mathbb{V}^{\dagger}\mathbb{M}_{L}\mathbb{U} = \mathrm{diag}(m_1,\ldots,m_n),
\end{equation}
with $V_{\alpha i} = \langle \alpha|i - \rangle $ and $U_{\alpha i} = \langle \alpha|i + \rangle $. Moreover, by virtue of Eq.~(\ref{symm}), $\mathbb{V}$ is expressed in terms of $\mathbb{U}$: $\mathbb{V} = \mathbb{U}^*$, 
so that
\begin{equation}  \label{symm2}
\mathbb{U}^{T}\mathbb{M}_{L}\mathbb{U} = \mathbb{V}^{T}\mathbb{M}_{R}\mathbb{V} = \mathrm{diag}(m_1,\ldots,m_n).
\end{equation}
By means of the $4n \times 4n$ matrix
$\hat{\mathbb{U}} = \mathbb{U}\Pi_{-} + \mathbb{V}\Pi _{+}$,
the mass term of the Lagrangian transforms into a diagonal matrix, 
\begin{equation} \label{4nx4n}
\hat{\mathbb{M}} \rightarrow \gamma ^{0}\hat{\mathbb{U}}^{\dagger}\gamma ^{0}\hat{\mathbb{M}}\hat{\mathbb{U}}
= \mathbb{V}^{\dagger} \mathbb{M}_{L} \mathbb{U}\Pi _{-}+ \mathbb{U}^{\dagger}\mathbb{M}_{R}\mathbb{V}\Pi _{+} = 
\hat{1} \otimes \mathrm{diag}\left(m_{1},...,m_{n}\right) ,
\end{equation}
where $\hat{1}$ is a unit matrix in the space of bispinors. Since  $\hat{\mathbb{U}}$ is real,
the neutrality of bispinors in the flavor basis is a necessary and sufficient condition
for the neutrality of bispinors diagonalizing the mass matrix. 
When the PMNS matrix $\mathbb{U}$ is employed, Eq (\ref{symm2}) yields a mass matrix $\mathbb{M}_{R}$ with non-zero phases on the diagonal. These phases are further eliminated by chiral rotation, as indicated below Eq.~(\ref{SM}), after which the mass matrix takes on the form (\ref{massmatrix}).

Using trace identities \cite{Kondratyuk:1992}, the characteristic polynomial (\ref{character}) for three neutrino flavors can be written as follows
\begin{eqnarray}
p(\lambda) = (\lambda - \lambda_0)^3 - \frac{1}{2}(\lambda - \lambda_0)\mathrm{Tr}[\mathbb{C}^2] - \frac{1}{3}\mathrm{Tr}[\mathbb{C}^3],
\end{eqnarray}
where
\begin{eqnarray} \label{expl1}
\mathbb{C} = \mathbb{M}_{\varsigma}\mathbb{M}_{- \varsigma} - \frac{1}{3}\mathrm{Tr}[\mathbb{M}_{\varsigma}\mathbb{M}_{- \varsigma}]
\end{eqnarray}
is a traceless Hermitian matrix, and
\begin{eqnarray} \label{expl2}
\lambda_0 = \frac{1}{3}\mathrm{Tr}[\mathbb{M}_{\varsigma}\mathbb{M}_{- \varsigma}]
\end{eqnarray}
is a real positive number that represents the average of neutrino masses squared. The coefficients of the characteristic polynomial are traces of the powers of $\mathbb{M}_{\varsigma}\mathbb{M}_{- \varsigma}$. 

The matrix $\mathbb{C}$  can be decomposed over the Gell--Mann--Okubo matrices: $\mathbb{C} = \omega_a\lambda^a$, where $\omega_a$ are real numbers ($a = 1,\ldots,8$). Suppose we reduce $\mathbb{C}$ to a diagonal form:
$\mathbb{C} = O \Lambda O^{\dagger}$, where $O$ is unitary. The diagonal traceless matrix $ \Lambda $ has the form:
$\Lambda = a\lambda^3 + b\lambda^8$. The eigenvalues of $\mathbb{C}$ are
$a + b/\sqrt{3}$, $- a + b/\sqrt{3}$, and $-2b/\sqrt{3}$.
For quantities $a$ and $b$, solving the cubic equation by the Vieta method, we find
\begin{eqnarray}
a &=& ~~ \sqrt{\frac{\mathrm{Tr}[\mathbb{C}^2]}{2}} \sin\Theta,  \label{expl3} \\
b &=& - \sqrt{\frac{\mathrm{Tr}[\mathbb{C}^2]}{2}}  \cos\Theta,  \label{expl4}
\end{eqnarray}
where
\begin{eqnarray} \label{expl5}
\Theta = \frac{1}{3}\arccos\left(
\frac{\sqrt{6}\mathrm{Tr}[\mathbb{C}^3]}{(\mathrm{Tr}[\mathbb{C}^2])^{3/2}}\right).
\end{eqnarray}
The masses of diagonal neutrinos are finally expressed in terms of $\mathbb{M}^\dagger \mathbb{M}$:
\begin{eqnarray}
m_1^2 &=& \lambda_0 + a + \frac{b}{\sqrt{3}}, \nonumber \\
m_2^2 &=& \lambda_0 - a + \frac{b}{\sqrt{3}}, \nonumber \\
m_3^2 &=& \lambda_0 - \frac{2b}{\sqrt{3}}.    \label{expl6}
\end{eqnarray}
The condition of $m_i^2$ to be real also follows from the inequality
$(\mathrm{Tr}[\mathbb{C}^2])^3 \geq 6(\mathrm{Tr}[\mathbb{C}^3])^2$,
which holds for all $3\times3$ traceless Hermitian matrices.
The eigenvalues of $\mathbb{M}_{R}\mathbb{M}_{L}$ coincide with the eigenvalues of $\mathbb{M}_{L}\mathbb{M}_{R}$.
In general, there are three mass eigenstates for three Majorana neutrinos. \textcolor{black}{The ordering of $m_i^2$ depends on the angle (\ref{expl5}).}

Using Sylvester's matrix theorem \cite{Roger:1991}, the propagator (\ref{SF2}) can be written in the form:
\begin{equation}
\hat{\mathbb{S}}_{F}(p)=\sum_{i\varsigma}\frac{1}{p^{2}-m_{i}^{2}}
 \left(\hat{p}+\mathbb{M}_{- \varsigma}\right) |i \varsigma \rangle \langle i\varsigma| \Pi_\varsigma. \label{LProj}
\end{equation}
The neutrino states are projected onto the eigenstates of $  {\mathbb{M}}_{\varsigma} {\mathbb{M}}_{ - \varsigma}$
with the use of the Frobenius covariants \cite{Roger:1991}:
\begin{eqnarray}
|i \varsigma \rangle \langle i \varsigma| \equiv \mathbb{F}_{i \varsigma} = \prod_{j \neq i}\frac{\mathbb{M}_{\varsigma}\mathbb{M}_{- \varsigma} - \lambda_j}{\lambda_i - \lambda_j}. \label{expl7}
\end{eqnarray}
The mass eigenstates with $\varsigma = \pm 1$ form complete bases in flavor space:
\begin{equation} \label{compl}
\sum_{i} |i \varsigma \rangle \langle i \varsigma| = 1.
\end{equation} 

To illustrate the statements made, consider

\textbf{Example 1.} Let the mass matrix be given by the equation%
\begin{equation}
\hat{\mathbb{M}}=\lambda^{3}+i\gamma_{5}\lambda^{6},
\end{equation}
where $\lambda^{a}$ are the Gell-Mann matrices. The components of $\hat{\mathbb{M}}$ are measured in mass units. The standard representation
for $\lambda^{a}$\ is assumed to be identical with the flavor basis, so that%
\begin{equation}
\mathbb{M}_{L}=\mathbb{M}_{R}^{\ast}=\lambda^{3}-i\lambda^{6}=\left\Vert
\begin{array}
[c]{lll}%
1 & 0 & 0\\
0 & -1 & -i\\
0 & -i & 0
\end{array}
\right\Vert .
\end{equation}
The eigenvalues and eigenvectors can readily be found using the MAPLE
symbolic computation software package \cite{MAPLE}: 
\begin{equation}
(m_{Li})^T=(m_{Ri}^*)^{T}=\left\Vert 
\begin{array}{c}
1 \\ 
-1/\eta  \\ 
-\eta 
\end{array}%
\right\Vert ,\ \ \ \ \tilde{W}_{\alpha i} 
= \ \left\Vert 
\begin{array}{ccc}
1 & 0 & 0 \\ 
0 & -i/\eta  & -i\eta  \\ 
0 & 1 & 1%
\end{array}%
\right\Vert ,  \label{part}
\end{equation}%
where $\eta =$ $e^{i\pi /3}$,$\ $the index $\alpha $ numbers the rows, the
index $i$ numbers the columns. 
After normalizing the eigenvectors of $\mathbb{M}_{L}$: 
\begin{equation}
\tilde{W}_{\alpha i}\rightarrow W_{\alpha i}=(\sum_{\beta }\tilde{W}_{\beta
i}^{2})^{-1/2}\tilde{W}_{\alpha i},
\end{equation}
the matrix $\mathbb{W}$ becomes a complex orthogonal matrix obeying 
\[
\sum_{\alpha } {W}_{\alpha i} {W}_{\alpha j}=\delta _{ij}
\] or, in matrix notation, $\mathbb{W}^{T}\mathbb{W}=1$. 
The mass matrix is converted to the diagonal form in two steps. On the vector space of dimension $4n$, we define first 
\begin{equation}
\hat{\mathbb{W}}= \mathbb{W} \Pi _{-}+\mathbb{W}^{\ast }\Pi _{+}.
\end{equation}
The action of $\hat{\mathbb{W}}$ on the mass matrix gives
\begin{equation}
\hat{\mathbb{M}} \rightarrow \gamma ^{0}\hat{\mathbb{W}}^{T}\gamma ^{0}\hat{\mathbb{M}}\hat{\mathbb{W}}
=  {\mathbb{W}}^{T}\mathbb{M}_{L}{\mathbb{W}} \Pi _{-}
+ {\mathbb{W}}^{\dag }\mathbb{M}_{R} {\mathbb{W}^*} \Pi _{+}
= \mathrm{diag}\left( e^{i\gamma _{5}\vartheta _{1}}|m_{L1}|,...,e^{i\gamma
_{5}\vartheta _{n}}|m_{Ln}|\right) ,
\end{equation}
where $e^{- 2i\vartheta _{i}}=m_{Li}/m_{Ri}$. 
At the second stage, the chiral rotations $\nu _{i}\rightarrow \exp (- i\gamma _{5}\vartheta _{i}/2)\nu
_{i}$ eliminate the phase factors. $\hat{{\mathbb{W}}}$ is a real matrix, therefore, it converts Majorana fields to Majorana fields. The drawback of ${\hat {\mathbb{W}}}$ is its non-unitarity. 
According to Eq.~(\ref{symm2}), a unitary matrix exists that converts $\mathbb{M}_{\pm}$ into its diagonal form (see also Refs. \cite{Bilenky:2018,Schechter:1980}).

The matrices included in the denominator of the neutrino propagator can be found to be
\begin{equation}
\mathbb{M}_{R}\mathbb{M}_{L}=\left(  \mathbb{M}_{L}\mathbb{M}_{R}\right)
^{\ast}=\left\Vert
\begin{array}
[c]{ccc}%
1 & 0 & 0\\
0 & 2 & i\\
0 & -i & 1
\end{array}
\right\Vert .\label{xxxxxx}%
\end{equation}
The corresponding eigenvalues and eigenvectors are equal 
\begin{equation}
(m_{i}^{2})^{T}=\left\Vert 
\begin{array}{c}
1 \\ 
2-g \\ 
1+g%
\end{array}%
\right\Vert ,\ \ \ \ \tilde{U}_{\alpha i} = \tilde{V}_{\alpha i}^* =\left\Vert 
\begin{array}{ccc}
1 & 0 & 0 \\ 
0 & -i/g & ig \\ 
0 & 1 & 1%
\end{array}%
\right\Vert ,  \label{example8}
\end{equation}%
where $g=(1+\sqrt{5})/2$. The normalization gives%
\[
\langle \alpha |i+\rangle =\langle \alpha |i-\rangle ^{\ast }=
(\sum_{\beta } |\tilde{U}_{\beta i}|^2)^{-1/2}\tilde{U}_{\alpha i}.
\]%
It can be verified that Eq.~(\ref{bases}) applied to $\langle \alpha
|i+\rangle $ yields vectors proportional to $\langle \alpha |i-\rangle $.
Calculation of the parameters entering Eqs. (\ref{expl2})--(\ref{expl5}%
) gives $\lambda _{0}=4/3$, $a=\frac{2}{\sqrt{3}}\cos (\frac{1}{3}\arccos (%
\frac{11}{16}))$, $b=-\frac{2}{\sqrt{3}}\sin (\frac{1}{3}\arccos (\frac{11}{%
16}))$. Using relations (\ref{expl6}) leads to values $m_{i}^{2}$ that
coincide with those given above.

\begin{table}[t]
\centering
\caption{Four pairs of the lepton currents at points $x$ and $y$, leading to neutrino oscillations of various types, are presented in the first two columns; the next column shows the time ordering, the last column shows the processes described by the lepton currents
\label{tab:table1}
}
\begin{tabular}{|c c c c|}
\hline
$j_{\mu}^{\textrm{CC}}(x)$ & $j_{\mu}^{\textrm{CC}}(y)$ & Time ordering & Process \\ 
\hline
$\bar{\ell}_{\beta }^{-}\gamma _{\mu }(1-\gamma _{5})\nu_{\beta }$ 
& $\bar{\nu}_{\alpha }\gamma _{\mu }(1-\gamma_{5})\ell _{\alpha }^{-}$ & 
\begin{tabular}{l}
$x^{0}-y^{0}>0$ \\ 
$x^{0}-y^{0}<0$
\end{tabular}
& 
\begin{tabular}{l}
$\nu _{\beta L }\leftarrow  \nu _{\alpha L }$ \\ 
$\nu _{\beta R }\rightarrow \nu _{\alpha R }$
\end{tabular}
\\ 
$\bar{\ell}_{\beta }^{-c}\gamma _{\mu }(1+\gamma _{5})
{\nu }_{\beta }$ & $\bar{\nu}_{\alpha }\gamma _{\mu}(1-\gamma _{5})
\ell _{\alpha }^{-}$ & 
\begin{tabular}{l}
$x^{0}-y^{0}>0$ \\ 
$x^{0}-y^{0}<0$
\end{tabular}
& 
\begin{tabular}{l}
$\nu _{\beta R }\leftarrow \nu _{\alpha L }$ \\ 
$\nu _{\beta L }\rightarrow \nu _{\alpha R }$
\end{tabular}
\\ 
$\bar{\ell}_{\beta }^{-}\gamma _{\mu }(1-\gamma _{5})\nu
_{\beta }$ & $\bar{\nu}_{\alpha } \gamma _{\mu }(1+\gamma _{5})\ell _{\alpha }^{-c}$ & 
\begin{tabular}{l}
$x^{0}-y^{0}>0$ \\ 
$x^{0}-y^{0}<0$
\end{tabular}
& 
\begin{tabular}{l}
$\nu _{\beta L }\leftarrow \nu _{\alpha R }$ \\ 
$\nu _{\beta R }\rightarrow \nu _{\alpha L }$
\end{tabular}
\\ 
$\bar{\ell}_{\beta }^{-c}\gamma _{\mu }(1+\gamma _{5}){%
\nu }_{\beta }$ & $\bar{\nu}_{\alpha } \gamma _{\mu }(1+\gamma _{5})\ell _{\alpha }^{-c}$ & 
\begin{tabular}{l}
$x^{0}-y^{0}>0$ \\ 
$x^{0}-y^{0}<0$
\end{tabular}
& 
\begin{tabular}{l}
$\nu _{\beta R }\leftarrow \nu _{\alpha R }$ \\ 
$\nu _{\beta L }\rightarrow \nu _{\alpha L }$
\end{tabular} \\
\hline
\end{tabular}
\end{table}

\vspace{4mm}

Integrating the time-like component of the four-momentum in Eq.~(\ref{SFcoord}), we find neutrino
propagator in the coordinate representation
\begin{eqnarray}
i\hat{\mathbb{S}}_F(x-y) &=& \theta(x^0 - y^0)\sum_{i\varsigma} \int \frac{d\mathbf{p}}{(2\pi)^3 2E_i(\mathbf{p})}
e^{-iE_i(\mathbf{p})(x^0 - y^0) + i\mathbf{p}(\mathbf{x} - \mathbf{y})}
 \left(\hat{p}_i+\mathbb{M}_{-\varsigma}\right)  \mathbb{F}_{i \varsigma} \Pi_\varsigma
 \nonumber \\
&-&\theta(y^0 - x^0)\sum_{i \varsigma} \int \frac{d\mathbf{p}}{(2\pi)^32E_i(\mathbf{p})}
e^{+iE_i(\mathbf{p})(x^0 - y^0) - i\mathbf{p}(\mathbf{x} - \mathbf{y})}
\left(\hat{p}_i-\mathbb{M}_{-\varsigma}\right) \mathbb{F}_{i \varsigma} \Pi_\varsigma,
\end{eqnarray}
where $\hat{p}_i = E_i(\mathbf{p})\gamma_0 - \mbox{\boldmath$\gamma$}\mathbf{p}$ and $E_i(\mathbf{p}) = \sqrt{m_i^2 + \mathbf{p}^2}$.

Massive neutrinos change chirality during oscillations. The right-handed antineutrinos take part in weak interaction. Vertices with right-handed neutrinos should be taken into account because Majorana neutrinos are neutral. The matrix elements shown in the first row of Table \ref{tab:table1} control the emission and absorption of neutrinos. The charged leptons are represented by ${\ell }_{\alpha }^{-}=(e^{-},\mu
^{-},\tau ^{-})$. 
The next rows of Table \ref{tab:table1} contain three more pairs of matrix elements produced by permuting fermionic fields. The oscillations of the left-handed neutrino created at point $y$ into the left-handed neutrino absorbed at point $x$ are described by the matrix elements in the first row of Table \ref{tab:table1}.
The oscillations of the right-handed neutrino generated at point $x$ into the right-handed neutrino absorbed at point $y$ occur with the reverse ordering of time $x^{0}-y^{0}<0$. 

It is not difficult to identify the component of the neutrino propagator that triggers each of the processes shown in the table because projection operators $\Pi_{\pm}$ enter the vertices. 
The component of $\hat{\mathbb{S}}_{F}(x-y)\Pi _{+}$ proportionate to $\hat{p}_{i}\Pi _{+}$ is suitable 
in the case of $\nu _{\beta L }\leftarrow \nu _{\alpha L }$. When there are oscillations $\nu _{\beta R}\rightarrow \nu _{\alpha L}$, the term is proportional to  $\mathbb{M}_{L}\Pi _{+}$ is relevant, and so on.
The vertices listed in the table only partially act independently. For instance, the transformation of the time inversion relates the process $\nu _{\beta L }\leftarrow \nu _{\alpha L }$  to the process $\nu _{\beta R }\rightarrow \nu _{\alpha R }$ in the second row: $\mathcal{T}\hat{A}(\nu _{\beta L }\leftarrow \nu _{\alpha L
})\mathcal{T}^{-1}=\hat{A}(\nu _{\beta R }\rightarrow \nu _{\mathrm{R%
}\alpha })$. In relation to processes $B^c \to A^c$, all processes $A \to B$ share this feature.

The component of the amplitudes, $\Pi _{\upsilon }i\hat{\mathbb{S}}%
_{F}(x-y)\Pi _{\varsigma }$ ($\upsilon , \varsigma = \pm 1$), that describes the neutrino propagation, 
is represented by a $4\times 4$ matrix in the space of bispinors of the charged
leptons. Neglecting the spread in the neutrino momenta, the neutrino component can be represented as follows
\begin{eqnarray}
\hat{A}(\nu _{\beta L }\leftarrow \nu _{\alpha L })=-\hat{A}%
(\nu _{\beta R }\rightarrow \nu _{\alpha R }) &\propto& -\sum_{i}e^{-iE_{i}(\mathbf{p})\tau }\langle \beta |i + \rangle \langle i + |\alpha \rangle \frac{\hat{p}_{i}}{%
E_{i}(\mathbf{p})}\Pi _{+} \label{osc10} \\
&=&-\sum_{i}e^{-iE_{i}(\mathbf{p})\tau }\langle \beta | \mathbb{F}_{i +} |\alpha \rangle  \frac{\hat{p}_{i}}{E_{i}(%
\mathbf{p})}\Pi _{+},  \nonumber
\end{eqnarray}
where $\tau =|x^{0}-y^{0}|$ and $|\alpha \rangle $ are the neutrino flavor states. 
The matrix element 
\begin{equation} \label{PMNS}
\langle \alpha |i + \rangle = U_{\alpha i} 
\end{equation}
can be recognized
as the PMNS matrix element. 
The ratio ${\hat{p}_{i}}/{E_{i}(\mathbf{p})}$ is sensitive to the neutrino
masses for $|\mathbf{p}|\lesssim m_{i}$, 
the exponential factor $\exp (-iE_{i}(\mathbf{p})\tau )$ is sensitive to the neutrino
masses in a broad range of $\mathbf{p}$ due to the macroscopic time factor $\tau$,
the Frobenius covariant (\ref{expl7}) entering the second line depends on the neutrino masses from the start,
while the other components of the
amplitude have purely kinematic origin and are the same for all types of
neutrinos. 

\textcolor{black}{
In recent decades, the theory of neutrino oscillations in a vacuum and in a
medium has been studied in many details
\cite{Wolfenstein:1978,Mikheev:1985,Kobzarev:1982,deSalas:2017kay,Bilenky:2018,Kovalenko:2023,Gribov:1969,Naumov:2020}. 
In the limit of large values of neutrino momenta, the kinematic factors of
reactions involving neutrinos are factorized, and an expression for the
probability of oscillations can be found:
\begin{equation}
P(\nu _{\beta L }\leftarrow \nu _{\alpha L
})=\sum_{ij}e^{-i(E_{i}(\mathbf{p})-E_{j}(\mathbf{p}))t}U_{i\beta }^{\ast }U_{i\alpha }U_{j\beta
}U_{j\alpha }^{\ast },
\end{equation}%
where $t$ is the oscillation time. It is assumed that the momentum $\mathbf{p}$ of neutrinos is fixed. 
This equation becomes
\begin{equation}
P(\nu _{\beta L }\leftarrow \nu _{\alpha L
})=\sum_{ij}e^{-i(E_{i}(\mathbf{p})-E_{j}(\mathbf{p}))t}\langle \beta |\mathbb{F}_{i+}|\alpha
\rangle \langle \alpha |\mathbb{F}_{j+}|\beta \rangle 
\end{equation}
once formulae (\ref{expl7}) and (\ref{PMNS}) are taken into account.
This equation is also a consequence of Eq.~(\ref{osc10}).
}

\textcolor{black}{
The probability has a similar form in the spatial picture with fixed energy $E$. The
exponential factor changes to 
\begin{equation}
\exp( -i(E_{i}(\mathbf{p}) - E_{j}(\mathbf{p}))t) \approx \exp(-i \frac{ m_{i}^{2} - m_{j}^{2} } {2|\mathbf{p}|} t) 
\rightarrow 
\exp( -i \frac{m_{i}^{2}-m_{j}^{2}}{2E} L),
\end{equation}
where $L$ is the distance between the source and the detector. 
A description of the complete kinematic structure of the oscillation amplitudes in the spatial 
representation can be found in ~\cite{Kovalenko:2023}.
}

\textcolor{black}{
In this paper, the analysis of oscillations is based on the plane-wave approximation \cite{Gribov:1969}. 
Despite the successful application of the plane-wave approximation, modeling processes using 
wave packets of the involved particles encompasses some challenges. 
Wave packets are employed in quantum field theory to accurately define asymptotic states, 
while scattering theory includes wave packets from the outset.
When the problem is reformulated using the wave packet techniques, the standard formulas for the probability 
of neutrino oscillations are reproduced; however, neutrino oscillations are expected to be suppressed 
at distances greater than the coherence and dispersion lengths typical for neutrinos arriving from galactic 
sources \cite{Naumov:2020}.
}

Chirality-changing transitions of neutrinos are of interest in the
context of double-$\beta $ decay. The process $\nu _{\beta R
}\leftarrow \nu _{\alpha L }$ is described by the currents shown in
the second raw of Table \ref{tab:table1}. The same matrix elements are responsible for the
process $\nu _{\beta R }\rightarrow \nu _{\alpha L }$. The
component of the amplitude associated with the neutrino propagation equals $%
\Pi _{+}i\hat{\mathbb{S}}_{F}(x-y)\Pi _{+}$, so we obtain 
\begin{eqnarray}
\hat{A}(\nu _{\beta R }\leftarrow \nu _{\alpha L })=-\hat{A}%
(\nu _{\beta R }\rightarrow \nu _{\alpha L }) &\propto
&-\sum_{i}e^{-iE_{i}(\mathbf{p})\tau }\ m_{i}\langle i +|\beta \rangle \langle
i+|\alpha \rangle  \frac{1}{E_{i}(\mathbf{%
p})}\Pi _{+}  \label{osc2} \\
&=&-\sum_{i}e^{-iE_{i}(\mathbf{p})\tau }\langle \beta | \mathbb{M}_{L} \mathbb{F}_{i +} |\alpha \rangle 
\frac{1}{E_{i}(\mathbf{p})}\Pi _{+}.  \nonumber
\end{eqnarray}%
Initially, under the sum sign, the expression $\langle \beta |\mathbb{M%
}_{L}|i+\rangle \langle i+|\alpha \rangle $ occurs. In order to obtain the
first line, we use the identities $\langle \beta |\mathbb{M%
}_{L}|i + \rangle $ = $m_{i}\langle \beta |i -\rangle $ = $m_{i}\langle \beta |i + \rangle
^{* }$ = $m_{i}\langle i +|\beta \rangle $, which also allow calculating
the mass matrix $\langle \beta |\mathbb{M}_{L}|\alpha \rangle $ in terms of
the neutrino masses and the mixing matrix: 
\begin{equation} \label{nice}
\langle \beta |\mathbb{M}_{L}|\alpha \rangle =\sum_{i}m_{i}\langle i+|\beta
\rangle \langle i+|\alpha \rangle .
\end{equation}
This equation is equivalent to Eq.~(\ref{symm2}). 
The mass matrix of the right chirality is determined by $\langle \beta |%
\mathbb{M}_{R}|\alpha \rangle =\langle \alpha |\mathbb{M}_{L}|\beta \rangle
^{\ast }$. 
We remark out that, as can be seen from Example 1, 
the eigenvalues of the matrix $\mathbb{M}_{L}$ do not match the masses $m_i$ 
obtained by diagonalizing the matrix $\mathbb{M}_{R}\mathbb{M}_{L}$.
Section IV gives further details of the double-$\beta $ decay
process.

Thus, the amplitude of processes associated with the emission and
absorption of Majorana neutrinos is expressed explicitly in terms of the
mass matrix using Eqs.~(\ref{expl1})--(\ref{expl6}) and (\ref{expl7}).

\subsection{Dirac neutrino}

The results can be generalized to the case of Dirac neutrinos. The Lagrangian for Dirac neutrinos has the form
\begin{equation} \label{LD}
\mathcal{L}_{D}= \bar{\nu}_{\alpha} i\hat{\nabla}\nu _{\alpha} - \bar{\nu}_{\alpha} \left( M_{\alpha \beta}+i\gamma _{5}N_{\alpha \beta}\right) \nu _{\beta},
\end{equation}
where $\nu _{\alpha}$ are the four-component complex spinors of Dirac neutrinos.
Hermiticity of the Lagrangian implies that the matrices $\mathbb{M}$ and
$\mathbb{N}$ are Hermitian, whereas $\mathbb{M}_{\pm } = \mathbb{M} \pm i \mathbb{N}$ are arbitrary complex matrices satisfying $\mathbb{M}_{+}^\dagger = \mathbb{M}_{-}$.
The mass matrix is defined by $\hat{\mathbb{M}} = \mathbb{M}+i\gamma_5 \mathbb{N}$.
The neutrino propagator has the form of Eq.~(\ref{SF2}). Because $\mathbb{M}_{R}\mathbb{M}_{L}$ is Hermitian, Eqs. (\ref{SFcoord})--(\ref{compl}) apply to Dirac neutrinos, except for (\ref{symm}) and (\ref{symm2}).

\textcolor{black}{In the  Weyl representation of two-component spinors $\nu _{\alpha L}$ and $\nu _{\alpha R}$, the Lagrangian (\ref{LD}) takes the form%
\begin{equation}
\mathcal{L}_{M}= \bar{\nu}_{\alpha L}i\hat{\nabla}\bar{\nu}_{\alpha L}
- \bar{\nu}_{\alpha L}M_{\alpha \beta}^{{D}} \nu _{\beta R} + \mathrm{h.c.},
\end{equation}%
where $M^{{D}}_{\alpha \beta} =  M_{\alpha \beta} + i N_{\alpha \beta}$ 
is a complex $n\times n$ matrix  $\mathbb{M}_{R}$.
}

The matrix $\mathbb{M}_{R}\mathbb{M}_{L}$ for Majorana and Dirac neutrinos is controlled by $n^2$ independent 
real parameters. 
Majorana neutrinos have $n(n+1)$ and Dirac neutrinos have $2n^2$ parameters 
in the matrix $\mathbb{M}_{L}$ (or $\mathbb{M}_{R}$). 
The diagonalization of $\mathbb{M}_{R}\mathbb{M}_{L}$ does not automatically imply that in the derived basis $\mathbb{M}_{L}$ (or $\mathbb{M}_{R}$) is diagonal.
The only result that follows from arguments used in the previous subsection is Eq.~(\ref{4nx4n}),
where in the case of Dirac neutrinos $\mathbb{U}$ and $\mathbb{V}$ are two distinct unitary matrices. 
\textcolor{black}{
The unitary matrices  $\mathbb{V}$ and $\mathbb{U}$ in Eq.~(\ref{VMU}) can be expressed as 
$\mathbb{V} = \mathbb{P}_1\mathbb{V}^{\prime}\mathbb{P}_2$ and $\mathbb{U} = \mathbb{P}_3\mathbb{U}^{\prime}\mathbb{P}_4$, respectively, where $\mathbb{P}_i = \mathrm{diag}(e^{i\chi_{i1}},\ldots,e^{i\chi_{in}})$. 
Without reducing generality, $\chi_{21} = \chi_{41} = 0$ can be used, resulting in the unitary matrices $\mathbb{V}^{\prime}$ and $\mathbb{U}^{\prime}$ being defined by $n^2 - 2n + 1$ independent parameters. 
Since the right-handed neutrinos do not participate in interactions, it is possible to proceed to their linear combination by a unitary transformation.
Rotating the right-handed neutrinos with $\mathbb{P}_3^{\dagger} \mathbb{V}$ removes  $n^2$ parameters. By phase rotations 
$\nu_{L} \to \mathbb{P}_4^{\dagger}  \nu_{L}$,
the $n - 1$ additional parameters can be removed. The number of independent parameters of the mass matrix becomes $n^2 - n + 1$. This result is the number of diagonal masses $m_i$ plus the number of parameters of $\mathbb{U}^{\prime}$. For $n= 3$ the mass matrix is determined by $9 - 3 + 1 = 7$ parameters.
}

To illustrate the similarities and differences between the Dirac and Majorana mass matrices, consider

\textbf{Example 2.} Let the mass matrix be given by the equation%
\begin{equation}
\hat{\mathbb{M}}=\lambda ^{3}+i\gamma _{5}\lambda ^{7}.
\end{equation}%
The components of $\hat{\mathbb{M}}$ are measured in mass units. The matrix is not symmetric, but it satisfies the condition of Dirac
self-conjugacy $\gamma ^{0}\hat{\mathbb{M}}^{\dag }\gamma ^{0}=\hat{\mathbb{M%
}}$. The left and right chiral projections of $\hat{\mathbb{M}}$ read%
\begin{equation}
\mathbb{M}_{L} =\lambda ^{3}-i\lambda ^{7}=\left\Vert 
\begin{array}{rrr}
1 & 0 & 0 \\ 
0 & -1 & 1 \\ 
0 & -1 & 0%
\end{array}%
\right\Vert ,\ \ \ \ \mathbb{M}_{R}=\lambda ^{3}+i\lambda ^{7}=\left\Vert 
\begin{array}{rrr}
1 & 0 & 0 \\ 
0 & -1 & -1 \\ 
0 & 1 & 0%
\end{array}%
\right\Vert .
\end{equation}%
Unlike Majorana neutrinos, $\mathbb{M}_{L}$ and $\mathbb{M}_{R}$ are not complex conjugate. The
eigenvalues and unnormalized eigenvectors can be found to be: 
\[
(m_{Li})^T = (m_{Ri})^T = \left\Vert 
\begin{array}{c}
1 \\ 
-1/\eta  \\ 
-\eta 
\end{array}%
\right\Vert ,\ \ \ \ \tilde{W}_{\alpha i}^L = \left\Vert 
\begin{array}{ccc}
1 & 0 & 0 \\ 
0 & 1/\eta  & \eta  \\ 
0 & 1 & 1%
\end{array}%
\right\Vert ,\ \ \ \ \tilde{W}_{\alpha i}^R = \left\Vert 
\begin{array}{ccc}
1 & 0 & 0 \\ 
0 & -1/\eta  & -\eta  \\ 
0 & 1 & 1%
\end{array}%
\right\Vert ,
\]%
where $\eta =$ $e^{i\pi /3}$. 
The matrices entering the denominators of the neutrino
propagator are equal to 
\begin{equation}
\mathbb{M}_{R}\mathbb{M}_{L}=\left\Vert 
\begin{array}{rrr}
1 & 0 & 0 \\ 
0 & 2 & -1 \\ 
0 & -1 & 1%
\end{array}%
\right\Vert ,\ \ \ \ \mathbb{M}_{L}\mathbb{M}_{R}=\left\Vert 
\begin{array}{rrr}
1 & 0 & 0 \\ 
0 & 2 & 1 \\ 
0 & 1 & 1%
\end{array}%
\right\Vert .  \label{yyyyyy}
\end{equation}%
The corresponding eigenvalues and eigenvectors become
\begin{equation}
(m_{i}^{2})^T=\left\Vert 
\begin{array}{c}
1 \\ 
2-g \\ 
1+g%
\end{array}%
\right\Vert ,\ \ \ \ \tilde{U}_{\alpha i} 
=\left\Vert 
\begin{array}{ccc}
1 & 0 & 0 \\ 
0 & 1/g & - g \\ 
0 & 1 & 1%
\end{array}%
\right\Vert ,\ \ \ \ \tilde{V}_{\alpha i} 
=\left\Vert 
\begin{array}{ccc}
1 & 0 & 0 \\ 
0 & -1/g & g \\ 
0 & 1 & 1%
\end{array}%
\right\Vert ,  \label{example8bis}
\end{equation}%
where $g=(1+\sqrt{5})/2$. Calculation of the parameters entering Eqs. (\ref{expl2})--(\ref{expl5}) gives $\lambda _{0}=4/3$, $a=\frac{2}{\sqrt{3}}%
\sin (\frac{1}{3}\arccos (\frac{11}{16}))$, $b=-\frac{2}{\sqrt{3}}\cos (%
\frac{1}{3}\arccos (\frac{11}{16}))$. Using the relations (\ref{expl6})
leads to values $m_{i}^{2}$ that coincide with those given above. A significant difference from the Majorana mass case is that two chiral projections require two mixing matrices, 
\begin{eqnarray}
\langle \alpha |i+\rangle &=& (\sum_{\beta } |\tilde{U}_{\beta i}|^2)^{-1/2}\tilde{U}_{\alpha i}, \\
\langle \alpha |i-\rangle &=& (\sum_{\beta } |\tilde{V}_{\beta i}|^2)^{-1/2}\tilde{V}_{\alpha i},
\end{eqnarray}
to diagonalize the propagator. The relationship  (\ref{symm}) for Dirac neutrinos is no longer valid, as can be seen from the example considered. 

\begin{table}[t]
\centering
\caption{First two columns show the lepton currents at points $x$ and $y$, leading to neutrino oscillations; the next column shows the time ordering, the last column shows the process corresponding to the lepton currents; the second and third rows show the lepton currents that produce oscillations of Majorana neutrinos, but not Dirac neutrinos
\label{tab:table2}
}
\begin{tabular}{|c c c c|}
\hline
$j_{\mu}^{\textrm{CC}}(x)$ & $j_{\mu}^{\textrm{CC}}(y)$ & Time ordering & Process \\ 
\hline
$\bar{\ell}_{\beta }^{-}\gamma _{\mu }(1-\gamma _{5})\nu
_{\beta }$ & $\bar{\nu}_{\alpha }\gamma _{\mu }(1-\gamma
_{5})\ell _{\alpha }^{-}$ & 
\begin{tabular}{l}
$x^{0}-y^{0}>0$ \\ 
$x^{0}-y^{0}<0$
\end{tabular}
& 
\begin{tabular}{c}
$\nu _{\beta L }\leftarrow \nu _{\alpha L }$ \\ 
$\left( \nu ^{c}\right) _{\beta R }\rightarrow \left( \nu ^{c}\right)
_{\alpha R }$
\end{tabular}
\\ 
$\bar{\nu}_{\beta } \gamma _{\mu }(1-\gamma _{5})\ell
_{\beta }^{-}$ 
& $\bar{\nu}_{\alpha }\gamma _{\mu}(1-\gamma _{5})\ell _{\alpha }^{-}$ & 
$x^{0}-y^{0}\gtrless 0$ 
& 
- \\ 
$\bar{\ell}_{\beta }^{-}\gamma _{\mu }(1-\gamma _{5}) \nu_{\beta }$ 
& 
$\bar{\ell}_{\alpha }^{-}\gamma _{\mu }(1-\gamma _{5}) \nu _{\alpha }$ & 
$x^{0}-y^{0} \gtrless 0$
& 
- \\ 
$\bar{\nu}_{\beta }\gamma _{\mu }(1-\gamma _{5})\ell_{\beta }^{-}$ 
& $\bar{\ell}_{\alpha }^{-}\gamma _{\mu }(1-\gamma_{5}) \nu _{\alpha }$ 
& 
\begin{tabular}{l}
$x^{0}-y^{0}>0$ \\ 
$x^{0}-y^{0}<0$
\end{tabular}
& 
\begin{tabular}{c}
$\left( \nu ^{c}\right) _{\beta R }\leftarrow \left( \nu ^{c}\right)
_{\alpha R } $\\ 
$\nu _{\beta L }\rightarrow \nu _{\alpha L }$
\end{tabular} 
\\
\hline
\end{tabular}
\end{table}

\vspace{4mm}

Table \ref{tab:table2} shows the possible types of neutrino oscillations caused by the weak lepton currents at the space--time points $x$ and $y$. The combinations are also shown that could lead to the oscillations of Dirac neutrinos in the presence of right currents. One part of the amplitudes is expressed through the other by means of time inversion. The transition amplitudes represented in the last row are obtained by permuting the arguments $x \leftrightarrow y$ of the amplitudes in the first row. The oscillation amplitudes corresponding to the first row are determined by Eqs.~(\ref{osc10}) with the right-handed Majorana neutrinos $\nu_{\alpha R}$ identified as the left-handed Dirac antineutrinos $(\nu_{\alpha L})^c$.

The conclusion that is crucial to this discussion is that the neutrino mass matrix explicitly enters the amplitude of the oscillations of Dirac neutrinos in general.

Dirac neutrino is a superposition of two Majorana neutrinos with masses that are equal in absolute values and opposite in signs (see, e.g., \cite{Krivoruchenko:1996}).
Although the case described here is essentially a modified version of the one previously discussed, it is  highlighted because of its significance in application. 
The mixed case of Majorana and Dirac neutrinos is reduced to Majorana neutrinos and is not discussed here.

\section{Neutrino oscillations in medium}
\renewcommand{\theequation}{III.\arabic{equation}}
\setcounter{equation}{0}

\begin{figure} [t] %
\begin{center}
\includegraphics[angle = -90, width=0.300\textwidth]{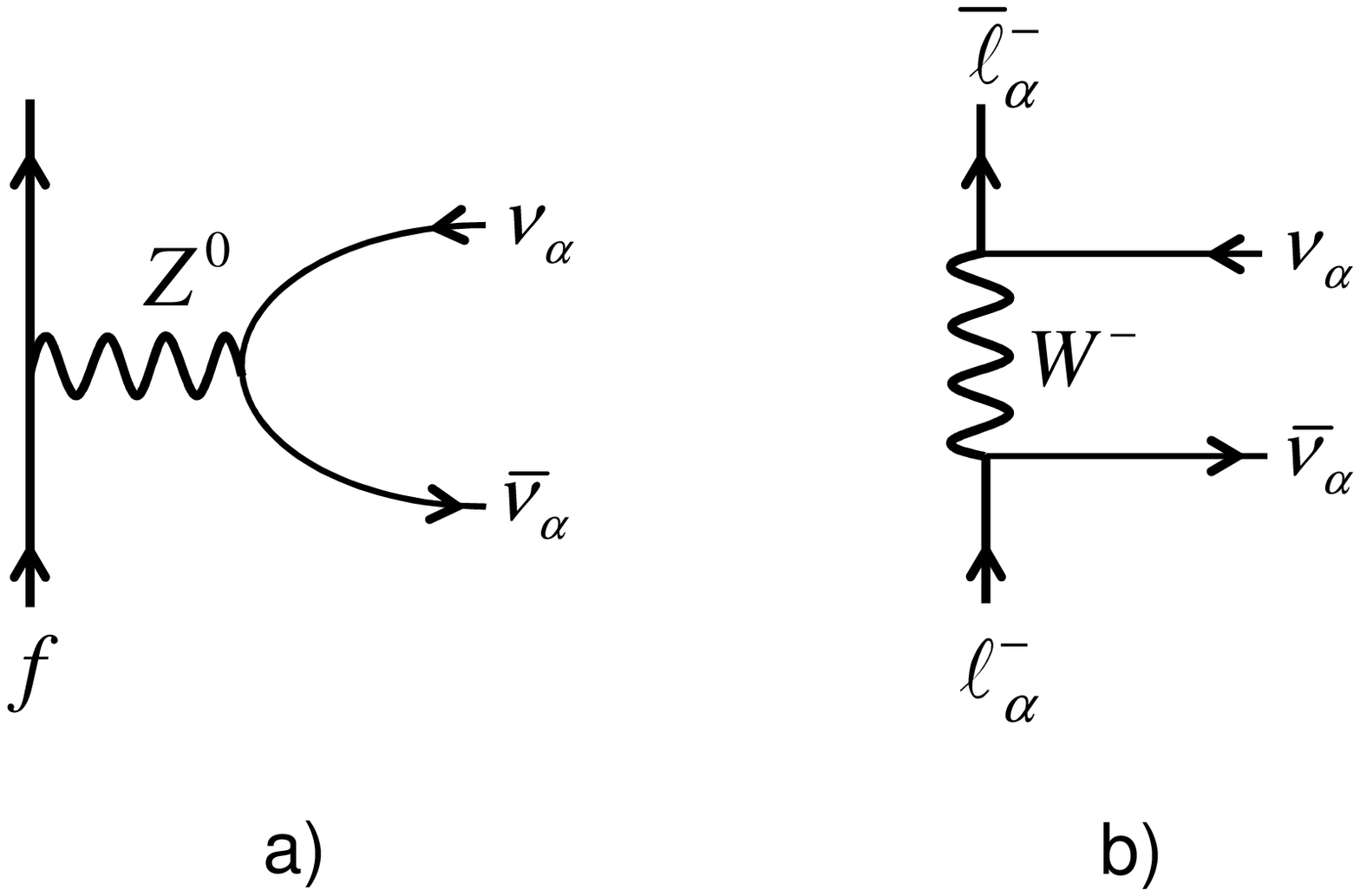}
\caption{
 Diagrams contributing to the mean-field potential of neutrino in matter.  $f$ is a Standard Model fermion, $\ell^-_{\alpha}$ is a charged lepton, and $\nu_{\alpha}$ are neutrinos with the $\alpha$ flavor. The $Z$ boson exchange in the $t$-channel $a$) produces a diagonal potential independent of neutrino flavor. Charged leptons and neutrinos interact additionally through the $s$-channel exchange of the $W^-$ boson, as shown in the diagram $b$). The individual components of the mean-field potential $b$) are determined by the partial lepton densities. The total potential is diagonal in the neutrino flavor.
The initial and final states of the medium particles are identical, since the mean-field potential is determined by the forward scattering amplitude at zero transmitted momentum.
}
\label{fig1}
\end{center}
\end{figure}

The electroweak mean-field potential created by the $Z^0$ boson and the exchange potential of the  $W^{\pm}$ boson generated by charged leptons in the medium have both an effect on neutrinos.
The mean field in the rest frame of matter has a time-like element: $(V_{\mu})_{\alpha \beta} = (V_{\alpha \beta},\mathbf{0)}$. The potential in the flavor basis is diagonal: $V_{\alpha \beta} = V_{(\alpha)}\delta_{\alpha \beta}$, as illustrated in Fig.~\ref{fig1}. Hereafter, $\mathbb{V}$ stands for a matrix containing the components $V_{\alpha \beta}$.

\subsection{Majorana neutrino}

The effective Lagrangian of Majorana neutrinos in the medium has the form
\begin{equation} \label{LM2}
\mathcal{L}_{M}^{\prime} = \frac{1}{2}\bar{\nu}_{\alpha} \left( i\hat{\nabla}\delta_{\alpha \beta} 
+ 
\hat{V}_{\alpha \beta}\gamma_5 \right) \nu _{\beta}- \frac{1}{2}%
\bar{\nu}_{\alpha} \left( M_{\alpha \beta}+i\gamma _{5}N_{\alpha \beta}\right) \nu _{\beta}.
\end{equation}
The mean-field potential $\hat{V}_{\alpha \beta} = (\gamma_0 V_{\alpha \beta},\mathbf{0})$
is determined by the forward scattering amplitude of neutrinos and depends on the composition of the substance.
The initial and final states of the scattering particles are identical.
Neutrino exchange with the medium particles via the $Z$ boson is independent of the neutrino flavor.
W$^{\pm}$ bosons contributes to  $\mathbb{V}$ through the $s$-channel exchange involving lepton component of the substance.
Figure 1 shows both types of diagrams, each generating a flavor-diagonal mean-field potential.
Consequently, the mean-field potential is generally represented as $ V_{\alpha \beta} =  V_{(\alpha)} \delta_{\alpha \beta}$.
For example, there are no $\tau^-$ mesons in interiors of neutron stars, therefore, the component of
$V_{(\alpha)}$ associated with the $\tau^-$ meson is zero.
The mean-field potential influences the left-handed neutrino component. The mean field is a symmetric matrix in flavor indices,
so its vector component disappears (cf. Eq.~(\ref{xsxsx})) and only its axial component remains in Eq.~(\ref{LM2}). In the Weyl basis
\begin{equation*}
\mathcal{L}_{M}^{\prime} = \bar{\nu}_{L \alpha} \left( i\hat{\nabla}\delta_{\alpha \beta} 
- \hat{V}_{\alpha \beta} \right) \nu _{L \beta} 
- \frac{1}{2}\bar{\nu}_{L \alpha}^c \left( M_{\alpha \beta} - i N_{\alpha \beta}\right) \nu _{L \beta}
- \frac{1}{2}\bar{\nu}_{L \alpha} \left( M_{\alpha \beta}+ i N_{\alpha \beta}\right) \nu _{L \beta}^c,
\end{equation*}
where $ \nu _{L \alpha}^c =  (\nu _{L \alpha})^c$ and $\bar{\nu}_{L \alpha}^c = \overline{\nu _{L\alpha }^{c}}$.

In the momentum representation, the Feynman propagator takes the form\footnote{
In Sect. III.A, we consider the scenario of $[\mathbb{V},\mathbb{M}_{\pm}] = 0$, which corresponds to either baryon matter partly free of leptons or a neutrino mass matrix partly diagonal in flavor, i.e.,
$V_{(\alpha)} = V_{(\beta)}$ and ${M}_{\pm \alpha \beta} \neq 0$ or 
$V_{(\alpha)} \neq V_{(\beta)}$ and ${M}_{\pm \alpha \beta} = 0$ for $\alpha \neq \beta$. 
The general case 
will be discussed elsewhere.
}
\begin{eqnarray}
\hat{\mathbb{S}}_F(p) &=& \frac{1}{\hat{p}-\hat{\mathbb{M}}+ 
\hat{\mathbb{V}}\gamma _{5}} \nonumber \\
&=&\left( \hat{p}+\hat{\mathbb{M}}^{\dagger}+ 
\hat{\mathbb{V}}\gamma _{5}\right) \frac{1}{p^{2}-\hat{\mathbb{M}}\hat{\mathbb{M}}^{\dagger}-
\hat{\mathbb{V}}^{2}+
\left( \hat{p}\hat{\mathbb{V}}-\hat{\mathbb{V}}\hat{p}\right) \gamma _{5}}.
\end{eqnarray}
Unlike the case discussed in Section II, the chiral projections do not eliminate $\gamma$-matrices completely from the denominator.
Given $\hat{p}\hat{\mathbb{V}}-\hat{\mathbb{V}}\hat{p} = 2\mathbb{V}\mbox{\boldmath$\alpha$}\mathbf{p}$ in the rest frame of matter, it is helpful to decompose $\hat{\mathbb{S}}_F(p)$ further using projections into states where the scalar product of the matrix $\mbox{\boldmath$\alpha$}$ and the momentum $\mathbf{p}$ equals $\pm |\mathbf{p}| $.
The corresponding projection operators are defined by $P_{\pm} = (1 \pm\mbox{\boldmath$\alpha$} \mathbf{n})/{2}$,
where $\mathbf{n} = \mathbf{p}/|\mathbf{p}|$, $P_{\sigma}$ commute with $\Pi_{\varsigma}$.
In the standard representation and in the Weyl representation of the $\gamma$-matrices, the relationship $\mbox{\boldmath$\alpha$} = \gamma_5 \otimes \mbox{\boldmath$\sigma$}$ holds true, so that  $\mbox{\boldmath$\alpha$}\mathbf{n}$ can be interpreted as chirality times helicity ($\sigma = \varsigma \eta$, $\eta$ is helicity).


The neutrino propagator is represented by sum of the four projections:
\begin{eqnarray}
\hat{\mathbb{S}}_F(p)  &=& \sum_{\varsigma \sigma} \left( \hat{p} + {\mathbb{M}}_{- {\varsigma}} + {\varsigma} 
\hat{\mathbb{V}} \right)
\frac{1}{p^{2}-{\mathbb{M}}_{{\varsigma}}{\mathbb{M}}_{- {\varsigma}}-
{\mathbb{V}}^{2}  + 2{\varsigma} \sigma \mathbb{V}|\mathbf{p}| }\Pi_{\varsigma} P_\sigma. \label{RP}
\end{eqnarray}
The denominator is a Hermitian matrix, so the eigenvalues are real. 
They are determined from the characteristic equation
\begin{eqnarray} \label{char}
\det \left\Vert \lambda - {\mathbb{M}}_{{\varsigma}}{\mathbb{M}}_{- {\varsigma}}-
{\mathbb{V}}^{2} + 2\eta \mathbb{V}|\mathbf{p}| \right\Vert = 0.
\end{eqnarray}
The value of $\lambda$ depends on the momentum. 
The complex conjugation of the characteristic equation permutes the matrices ${\mathbb{M}}_{R} $ and ${\mathbb{M}}_{L}$, so the poles of the expression (\ref{RP}) depend on the helicity $\eta$ only. 
The dispersion law takes the form
\begin{eqnarray} \label{ds}
E_{i\eta}(\mathbf{p}) = \sqrt{\lambda_{i\eta} + \mathbf{p}^2},
\end{eqnarray}
where $\lambda_{i\eta}$ is one of the solutions.
Entering the determinant (\ref{char}) is a $3\times3$ matrix. This suggests that we are 
dealing with a cubic equation.
To find the roots, one can use the Vieta method, which has already been employed to analyze neutrino oscillations in vacuum.

The procedure is equivalent to that used in Section II.
We redefine the matrix $\mathbb{C}$ and the parameter $\lambda_0$:
\begin{eqnarray}
\mathbb{C} &=& \mathbb{M}_{\varsigma}\mathbb{M}_{- \varsigma} + 
{\mathbb{V}}^{2} - 2\eta \mathbb{V}|\mathbf{p}| 
- \frac{1}{3}\mathrm{Tr}\left[ \mathbb{M}_{\varsigma}\mathbb{M}_{- \varsigma} + 
{\mathbb{V}}^{2} - 2\eta \mathbb{V}|\mathbf{p}| \right], \\
\lambda_0  &=& \frac{1}{3}\mathrm{Tr}\left[\mathbb{M}_{\varsigma}\mathbb{M}_{- \varsigma} + 
{\mathbb{V}}^{2} - 2 \eta \mathbb{V}|\mathbf{p}| \right].
\end{eqnarray}
The formulas (\ref{expl1})--(\ref{expl6}) and (\ref{expl7}) allow to find 
the eigenvalues $\lambda_{i\eta}$ and the in-medium dispersion law.

The eigenstates determined by the equation
\begin{eqnarray} \label{eigenstate1}
\left( \mathbb{M}_{\varsigma}\mathbb{M}_{- \varsigma} + 
{\mathbb{V}}^{2} - 2 \eta \mathbb{V}|\mathbf{p}| \right) | i \varsigma \eta \rangle = \lambda_{i\eta} | i \varsigma \eta \rangle
\end{eqnarray}
depend on the momentum. The values of $E_{i\eta}(\mathbf{p})$ are real, since
\[
E_{i\eta}^2(\mathbf{p}) = \langle {i \varsigma  \eta}| \left( (\mathbb{M}_{- \varsigma})^\dagger \mathbb{M}_{- \varsigma} 
+ \left(|\mathbf{p}| - 
\eta \mathbb{V}  \right)^2 \right) | i \varsigma \eta \rangle \geq 0.
\]
The eigenvalues of Eq.~(\ref{eigenstate1}) are real and coincide for $\varsigma = \pm 1$,
while the coordinates of  $| i\pm \eta \rangle$ in the flavor basis are related by the equation 
\begin{equation} \label{symmetry}
\langle \alpha | i \varsigma \eta \rangle^* = \langle \alpha |  i - \varsigma  \eta \rangle.    
\end{equation}

In the simplest case of the mean-field potential proportional to the unit matrix,
which occurs when the diagram in Fig. 1 b) is turned off,
the momentum dependence of the dispersion law is quite transparent.
Assuming $V_{(\alpha)} = V$ for $\alpha = 1,2,3$, we find
\begin{eqnarray} \label{dlm}
E_{i\eta}(\mathbf{p}) =  \sqrt{m_i^2 + \left(|\mathbf{p}| - 
\eta V \right)^2},
\end{eqnarray}
where $m_{i}^2$ are the eigenvalues of $\mathbb{M}_{\varsigma}\mathbb{M}_{- \varsigma}$. 
Since $\mathbb{V}$ is commutative with  $\mathbb{M}_\varsigma$, the dependence on $\eta$ disappears from $| i\varsigma \eta \rangle $.

The equation for eigenvalues in an arbitrary potential admits a simple solution for neutrinos with a large momentum. In the left side of Eq.~(\ref{eigenstate1}), the leading term is $2 \eta \mathbb{V}|\mathbf{p}|$ with $\eta  = \pm 1$. Since the potential is diagonal in flavor, the eigenstates are also diagonal in flavor. The first two terms on the left side of the equations give a correction to $ \lambda_{i\eta }$.
In the leading order, the eigenvalues, the dispersion law, and the eigenstates take the form
\begin{eqnarray}
\lambda_{\alpha \eta } &=& - 2 \eta V_{\alpha} |\mathbf{p}|  + m_{\alpha}^2   + 
{V}_{\alpha}^{2} + \ldots, \label{major1}\\
E_{\alpha \eta} &=& |\mathbf{p}| -  
\eta V_{\alpha} + \frac{ m_{\alpha}^2}{2 |\mathbf{p}|} + \ldots, \label{major2} \\
|\alpha \varsigma \eta \rangle &=& |\alpha \rangle 
  + \ldots, \label{major3}
\end{eqnarray}
where $m_{\alpha}^2 = \langle \alpha | \mathbb{M}_{\varsigma}\mathbb{M}_{-\varsigma} | \alpha \rangle$. In the limit of $|\mathbf{p}| \to \infty$, the oscillations are expected to disappear for $V_{(\beta)} \neq V_{(\alpha)}$.
The degeneracy of the potential $V_{(\beta)} \approx V_{(\alpha)}$ for any pair of $\alpha \ne \beta $ restores the mixing
\cite{Wolfenstein:1978,Mikheev:1985}.
The mixing matrix (\ref{PMNS}) is split in the medium into two components depending on the parameter $\eta$:
\begin{equation} \label{spliu}
U^{\dagger}_{i\alpha} \to \langle i+|\alpha \varsigma \eta \rangle =  
U^{\dagger}_{i\alpha} 
+ \ldots
\end{equation}
The correction 
can be significant for low values of $|\mathbf{p}|$ or in the presence of degeneracy $V_{(\beta)} \approx V_{(\alpha)}$.

Integrating the time-like component of four-momentum in the coordinate representation, we obtain
\begin{eqnarray}
i\hat{\mathbb{S}}_{F}(x-y)&=&\theta (x^{0}-y^{0})\sum_{i\varsigma \eta}\int 
\frac{d\mathbf{p}}{(2\pi )^{3}2E_{i\eta}(\mathbf{p})}
e^{-iE_{i\eta}(\mathbf{p})(x^{0}-y^{0})+i\mathbf{p}(\mathbf{x}-\mathbf{y})} \label{SFmed} \\
&&~~~~~~~~~~~~~~~~~~\times \left( \hat{p}_{i\eta}+{\mathbb{M}}_{- \varsigma} +
\varsigma 
\hat{\mathbb{V}}\right)
|i\varsigma \eta \rangle \langle i\varsigma \eta |\Pi _{\varsigma}
 P_{\sigma } \nonumber \\
&-&\theta (y^{0}-x^{0})\sum_{i\varsigma \eta}\int 
\frac{d\mathbf{p}}{(2\pi )^{3}2E_{i\eta}(\mathbf{p})}
e^{+iE_{i\eta}(\mathbf{p})(x^{0}-y^{0})-i\mathbf{p}(\mathbf{x}-\mathbf{y})} \nonumber  \\
&&~~~~~~~~~~~~~~~~~~\times \left( \hat{p}_{i\eta}-{\mathbb{M}}_{-\varsigma} - 
\varsigma 
\hat{\mathbb{V}}\right)
|i\varsigma \eta \rangle \langle i \varsigma \eta |\Pi _{\varsigma} P_{\sigma }, \nonumber
\end{eqnarray}
where $\sigma = \varsigma \eta$.

The Frobenius covariants are given by
\begin{eqnarray}
|i \varsigma \eta \rangle \langle i \varsigma \eta | \equiv \mathbb{F}_{i \varsigma \eta} = \prod_{j \neq i}\frac{\mathbb{M}_{\varsigma} \mathbb{M}_{- \varsigma} + 
{\mathbb{V}}^{2} - 2 \eta \mathbb{V}|\mathbf{p}|  - \lambda_{j \eta}}{\lambda_{i \eta} - \lambda_{j \eta}}. \label{expl12}
\end{eqnarray}

Like in the previous section we consider first neutrino oscillations
$\nu_{\beta L} \leftarrow\nu_{\alpha L}$,
described by the current $j_{\mu}^{\textrm{CC}} = \bar{\nu}_{\alpha} \gamma_{\mu}(1-\gamma_{5})\ell_{\alpha}^{-}$
at the neutrino creation point $y$ and by the current 
$j_{\mu}^{\textrm{CC}} = \bar{\ell}_{\beta}^{-}\gamma_{\mu}(1-\gamma_{5}) \nu_{\beta}$ at the neutrino
absorption point $x$. A process is also
possible in which the right-handed neutrino $\nu_{\alpha R}$ is created at the point
$x$ and propagates to the point $y$ without chirality flip, where it is absorbed ($x^{0}-y^{0}<0$).
Neglecting the neutrino momentum spread, the neutrino component of the amplitude,
$\Pi_{-}i\hat{\mathbb{S}}_{F}(x-y)\Pi_{+}$, becomes
\begin{eqnarray} \label{osc1}
\hat{A}(\nu_{\beta L} \leftarrow\nu_{\alpha L}) &\propto& - \sum_{i\eta}
e^{-iE_{i\eta}(\mathbf{p})\tau}\langle\beta| \left(  \hat{p}_{i\eta} + 
\hat{\mathbb{V}}  \right) 
\mathbb{F}_{i \varsigma \eta} |\alpha\rangle 
 \frac{1}{E_{i\eta}(\mathbf{p})}
\Pi_{\varsigma} P_{\eta}, \\
\hat{A}(\nu_{\beta R} \rightarrow\nu_{\alpha R}) &\propto& ~~~\sum_{i\eta}
e^{-iE_{i\eta}(\mathbf{p})\tau}\langle\beta| \left(  \hat{p}_{i\eta} - 
\hat{\mathbb{V}}  \right) 
\mathbb{F}_{i \varsigma \eta} |\alpha\rangle 
 \frac{1}{E_{i\eta}(\mathbf{p})} \Pi_{\varsigma} P_{\eta},
\end{eqnarray}
where $\varsigma = +1$.

The mixing matrices with chirality $\varsigma = \pm 1$ are connected by Eq.~(\ref{symmetry}), however to characterize the mixing of neutrinos with helicity $\eta = \pm 1$, two distinct matrices $\langle \alpha|i+ +  \rangle$ and $\langle \alpha|i+-  \rangle$ are required.
In the limit of vanishing matter density, the matrices  $\langle \alpha|i+ \pm  \rangle$ become the PMNS matrix.

The process $\nu_{\beta R}\leftarrow \nu_{\alpha L}$ with chirality flip
is associated with the lepton current
$j_{\mu}^{\textrm{CC}} = \bar{\nu}_{\alpha} \gamma_{\mu}(1-\gamma_{5})\ell_{\alpha}^{-}$ 
at the creation
point $y$ and by the lepton current $j_{\mu}^{\textrm{CC}} = \bar{\nu}_{\beta}  \gamma_{\mu}(1-\gamma_{5})\ell_{\beta}^{-}$
at the absorption point $x$. The process $\nu_{\beta R}\rightarrow
\nu_{\alpha L}$ is described for $x^{0}-y^{0}>0$ by the same currents.
The component of the amplitudes responsible for chirality-flip propagation of neutrinos equals
$\Pi_{+}i\hat{\mathbb{S}}_{F}(x-y)\Pi_{+}$, so we obtain
\begin{equation} \label{osc22}
\hat{A}(\nu_{\beta R} \leftarrow\nu_{\alpha L}) = -
\hat{A}(\nu_{\beta R} \rightarrow\nu_{\alpha L})
\propto - \sum_{i\eta } e^{-iE_{i\eta}(\mathbf{p})\tau} 
\langle \beta| \mathbb{M}_{L} \mathbb{F}_{i \varsigma \eta} | \alpha\rangle
e^{-i\phi_\alpha - i\phi_\beta} \frac{1}{E_{i\eta}(\mathbf{p})}\Pi_{\varsigma}P_{\eta},
\end{equation}
where $\varsigma = +1$. The explicit dependence on the neutrino mass matrix is obvious. 
The arguments given after Eq.~(\ref{osc2}) prove 
the equation $\langle \beta| \mathbb{M}_L |i\varsigma \eta \rangle = m_{i} \langle i\varsigma \eta | \beta \rangle$
for 
$\mathbb{V}$ commutative with $\mathbb{M}_{\pm}$.

\subsection{Dirac neutrino}

The in-medium Lagrangian for Dirac neutrinos is written as
\begin{equation} \label{LM2bis}
\mathcal{L}_{D}^{\prime} = \bar{\nu}_{\alpha} \left( i\hat{\nabla}\delta_{\alpha \beta} - \hat{V}_{\alpha \beta}\Pi_- \right) \nu _{\beta}
- \bar{\nu}_{\alpha} \left( M_{\alpha \beta}+i\gamma _{5}N_{\alpha \beta}\right) \nu _{\beta}.
\end{equation}
The mean-field term $\hat{V}_{\alpha \beta}\Pi_-$ distinguishes it from the vacuum Lagrangian, $\mathcal{L}_{D}$.
The mass matrices $\mathbb{M}$ and $\mathbb{N}$ are Hermitian. The Feynman propagator has the following form in the momentum representation:
\begin{eqnarray}  \label{propdm}
\hat{\mathbb{S}}_{F}(p) &=&\frac{1}{\hat{p}-\hat{\mathbb{M}}-\hat{\mathbb{V}}\Pi _{-}}.
\end{eqnarray}

\begin{figure} [t] %
\begin{center}
\includegraphics[angle = 0, width=0.382\textwidth]{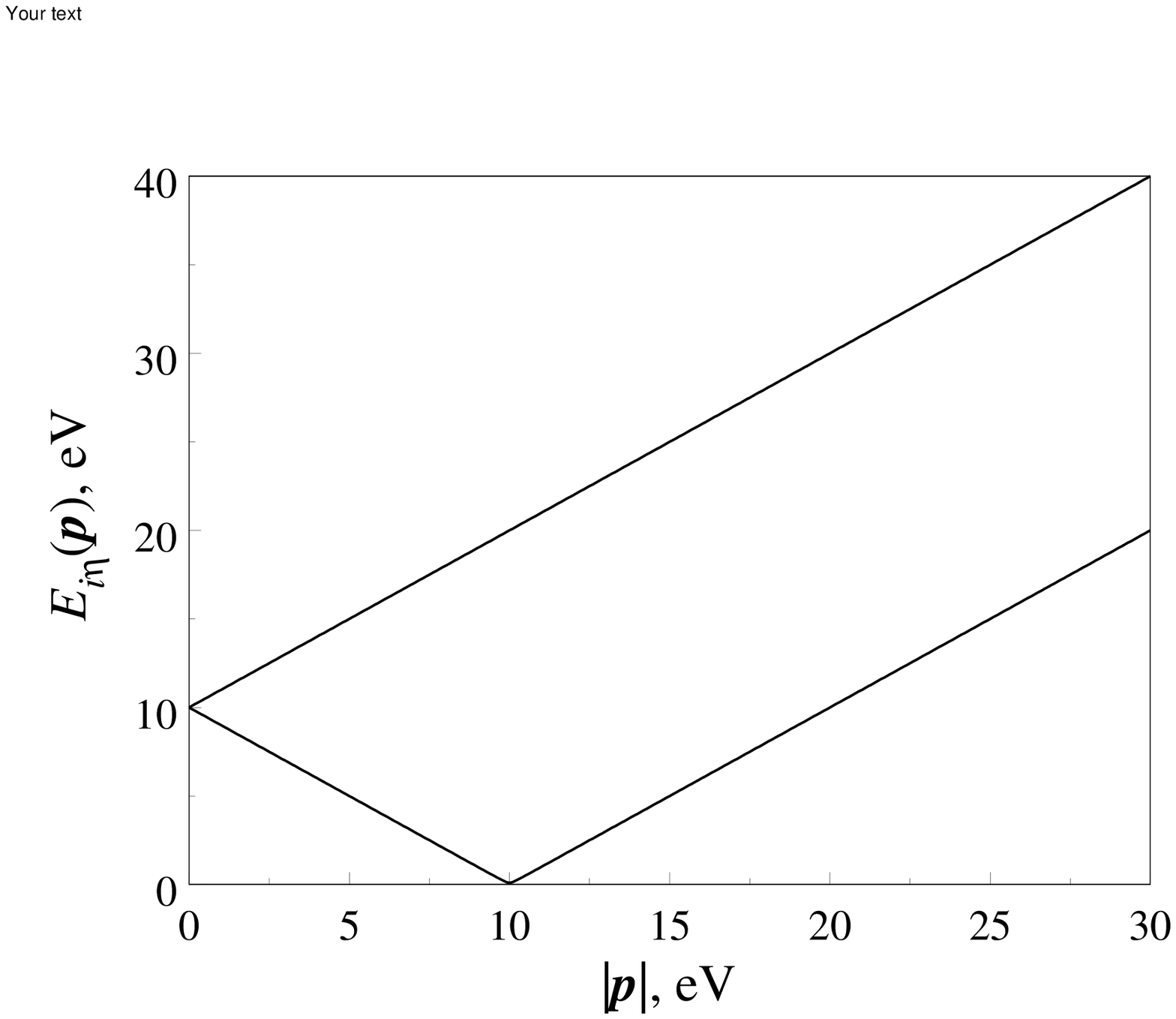}
\caption{
Two branches of dispersion law for Majorana neutrinos in a medium with a mean-field potential of $V = - 10$ eV, typical for nuclear matter at saturation,
and a neutrino mass of $m_i = 0.1$ eV.
The upper and lower lines correspond to neutrinos with left and right helicity $\eta = \mp 1$, respectively. 
The neutrino energy takes a minimum value $E_{i+}(\mathbf{p}) = m_i$ at $|\mathbf{p}| = 10$ eV.
The dispersion law of positive-energy Dirac neutrinos takes the same form for  $V = - 20$ eV
and is shifted along the vertical axis accordingly.
}
\label{fig2}
\end{center}
\end{figure}

\textbf{Example 3.} Given that $\hat{\mathbb{V}}$ is determined by time-like component of the four-vector,
the substitutions $E \to E - \mathbb{V}$ and $\mathbb{V} \to \mathbb{V}/2$ reduce the problem to the case of Majorana neutrinos provided $\mathbb{V}$ is proportional the unit matrix. In such a case, the dispersion law reads
\begin{equation} \label{osc6}
{E}_{i\eta}^{D\pm}(\mathbf{p}) = \frac{1}{2}V \pm E_{i\eta}(\mathbf{p}),
\end{equation}
where $E_{i\eta}(\mathbf{p})$ simplifies to (\ref{dlm}) with $V$ replaced by $V/2$.
In a neutron star, the mean-field potential generated by neutrons is $V \sim - 20$ eV (see, e.g., \cite{Krivoruchenko:2022}).
At zero temperature and for $V/2 + m_i < 0$ the potential well is filled first by Dirac neutrinos 
with a negative energy at $|\mathbf{p}| \sim |V|/2$,
then states with higher energy are filled.
At a low density of neutrinos, the design of the Fermi sphere turns out to be quite unusual:
the neutrino states form in the momentum space a hollow sphere with a radius close to $|V|/2$.
Figure \ref{fig2} depicts the dispersion law for the Majorana neutrino for $V = - 10$ eV. The positive-energy Dirac neutrino dispersion law is similar, but it is displaced vertically and has a modified scale of the horizontal axis.

The BCS mechanism causes a gap to appear in the neutrino excitation spectrum at ultra-low temperatures \cite{Kapusta:2004}. 
The pairing of Dirac neutrinos is caused by the exchange of the Higgs boson, which generates an attraction between 
fermions.
In the case under consideration, the BCS mechanism is applicable to the right--helicity neutrinos adjacent to both the outer surface of the hollow Fermi sphere and the inner.
At the temperature of old neutron stars of the order of 100 eV, condensation has not yet formed,
so the appearance of superfluidity can be expected only in the very distant future as a result of further cooling of neutron stars.
The right--helicity neutrinos contain an admixture of the left--chiral neutrinos, which interact in a repulsive manner due to the $Z^0$--boson exchange. The admixture of the left-chiral states is suppressed as $m_i/ E_{i+}(\mathbf{p})$, so a sufficiently small neutrino mass is required
to ensure the dominance of attraction due to the Higgs boson exchange.

The bottom curve of the dispersion law reaches the border of the continuum of states with negative energy at a depth of the potential well of $- 4m_i$.  This leads to the creation of neutrino-antineutrino pairs from vacuum and filling of levels with energy $E_i^{D+}(\mathbf{p}) < - m_i$. 
\vspace{4mm}

The technique for determining the dispersion law in a general scenario for in-medium Dirac neutrinos is more involved.
We decompose first the neutrino propagator (\ref{propdm}) in a series by powers of the
potential. 
The chirality-state projections of each of the sum terms can be simplified
by using Eq.~(\ref{SF2}). The summation of the outcome results in
\begin{eqnarray}
\hat{\mathbb{S}}_{F}(p)
&=&\left( \hat{p}+\mathbb{M}_{L}\right) \frac{1}{p^{2}-\mathbb{M}_{R}\mathbb{%
M}_{L}-\hat{\mathbb{V}}\hat{p}}\Pi _{+}   \nonumber \\
&+&\left( \hat{p}+\mathbb{M}_{R}\right) \frac{1}{p^{2}-\mathbb{M}_{L}\mathbb{%
M}_{R}}\Pi _{-}+\left( \hat{p}+\mathbb{M}_{L}\right) 
\frac{1}{p^{2}-\mathbb{M}_{R}\mathbb{M}_{L}-\hat{\mathbb{V}}\hat{p}}
\hat{\mathbb{V}}\mathbb{M}_{R}%
\frac{1}{p^{2}-\mathbb{M}_{L}\mathbb{M}_{R}}\Pi _{-}.  \label{simplify bis}
\end{eqnarray}
The second line, which refers to $\hat{\mathbb{S}}_{F}(p)\Pi_- $, has two types of poles. The first corresponds to a neutrino in a medium, while the second to a neutrino in a vacuum.
The second type pole is inactive.
To demonstrate this, we employ the chirality-state projectors to act on $\hat{\mathbb{S}}_{F}(p)\Pi_-$ from the left. Basic algebra gives
\begin{eqnarray}
\hat{\mathbb{S}}_{F}(p)\Pi _{-}
&=&\mathbb{M}_{L}
\frac{1}{p^{2}-\mathbb{M}_{R}\mathbb{M}_{L}-\hat{\mathbb{V}}\hat{p}}
\left( \hat{p}-\hat{\mathbb{V}}\right) \mathbb{M}_{L}^{-1}\Pi _{-} \nonumber \\ 
&+&
\frac{1}{p^{2}-\mathbb{M}_{R}\mathbb{M}_{L}- \hat{p} \hat{\mathbb{V}}}
\mathbb{M}_{R}\Pi _{-}. \label{simplify4}
\end{eqnarray}%
The transformations are carried out under the assumptions that the matrix  $\mathbb{V}$ only has a time-like component,
all matrices are non-singular, and the matrix  $\mathbb{V}$ generally does not commute with the mass matrix.
Since the poles of the free neutrino propagator vanish, the vacuum poles of the expression (\ref{simplify bis}) are fictitious. With the substitutions $E \rightarrow  E  - \mathbb{V}$ and $\mathbb{V} \to \mathbb{V}/2$, the propagator poles for Majorana neutrino and Dirac neutrino formally coincide.

The denominators of propagators 
turn into $n\times n$ matrices in the flavor space using projectors $P_{\sigma}$. 
Using 
the commutativity of $P_{\sigma}$ and $\Pi_{\varsigma}$, 
the  relationships $\hat{\mathbb{V}}\hat{p}P_{\sigma}=\mathbb{V}%
(E-\sigma|\mathbf{p}|)P_{\sigma}$, $P_{\sigma}\gamma_{0}=\gamma
_{0}P_{-\sigma}$ and $( \hat{p}-\hat{\mathbb{V}})P_{\sigma} = P_{- \sigma}( \hat{p}-\hat{\mathbb{V}})$,
the denominators of $\hat{\mathbb{S}}_{F}(p)\Pi_{\varsigma}P_{\sigma}$ 
can be transformed into $p^{2}-\mathbb{M}_{R}\mathbb{M}_{L}-\mathbb{V}(E-\eta|\mathbf{p}|)$. 
The excitation spectrum is determined thereby from the equation:
\begin{equation} 
\left(  \mathbb{M}_{\varsigma}\mathbb{M}_{- \varsigma}+\mathbb{V}(E-\eta
|\mathbf{p}|)\right)  |i\sigma\varsigma\rangle=\lambda_{i \eta}(E)|i\varsigma \eta \rangle.
\label{eigenvalueq}%
\end{equation}
It is useful to compare this equation to Eq.~(\ref{eigenstate1}). 

The characteristic equation of Dirac neutrinos differs from that of Majorana neutrinos in terms depending on the potential: 
\begin{eqnarray} \label{char2}
\det \left\Vert \lambda - {\mathbb{M}}_{{\varsigma}}{\mathbb{M}}_{- \varsigma} 
- \mathbb{V}( E - \eta |\mathbf{p}|) \right\Vert = 0.
\end{eqnarray}
For completeness, we also give the expressions of the matrix $\mathbb{C}$ and the parameter $\lambda_0$: 
\begin{eqnarray}
\mathbb{C} &=& 
\mathbb{M}_{\varsigma} \mathbb{M}_{- \varsigma} + \mathbb{V}( E - \eta |\mathbf{p}|)
- \frac{1}{3}\mathrm{Tr}\left[ \mathbb{M}_{\varsigma} \mathbb{M}_{- \varsigma} + \mathbb{V}( E - \eta |\mathbf{p}|) \right], \\ 
\lambda_0  &=& \frac{1}{3}\mathrm{Tr}\left[ \mathbb{M}_{\varsigma} \mathbb{M}_{- \varsigma} + \mathbb{V}( E - \eta |\mathbf{p}|) \right].
\end{eqnarray}
The relations (\ref{expl3} )--(\ref{expl5}) with $n=3$ are fully applicable for Dirac neutrinos in matter, in the relations (\ref{expl6}) it is necessary to replace $m_i^2 \to \lambda_{i \eta}(E)$. Unlike the Majorana
neutrino, $\lambda_{i\eta}(E)$ depends on the
energy. The characteristic equation, therefore, is a polynomial of degree $6$
with respect to the energy, and for a matrix $\mathbb{V}$ of general
form it is not solved in radicals. 

The Frobenius covariants read:
\begin{eqnarray}
|i \varsigma \eta \rangle \langle i \varsigma \eta | \equiv \mathbb{F}_{i \varsigma \eta}(E,\mathbf{p}) = 
\prod_{j \neq i}\frac{ \mathbb{M}_{\varsigma}\mathbb{M}_{- \varsigma}+\mathbb{V}(E - \eta
|\mathbf{p}|)  - \lambda_{j \eta}(E) }{\lambda_{i \eta}(E) - \lambda_{j \eta}(E)}. \label{expl15}
\end{eqnarray}

Let us consider the case of
large momenta. The dominant term $\mathbb{V}(-\eta
|\mathbf{p}|)$ in Eq. (\ref{eigenvalueq}) if not entirely compensated by the term $\mathbb{V} E$ 
implies that the zeroth approximation is formed by the flavor states. 
In the leading order, we obtain
\begin{align}
\lambda_{\alpha\eta}(E_{\alpha\eta}^{D+})  & = (1 -\eta) V_{\alpha}|\mathbf{p}| + m_{\alpha}^{2}+ 
\frac{1}{2}(1 -\eta)V_{\alpha}^2 +\ldots,\label{largepd1}\\
E_{\alpha\eta}^{D+}  & = |\mathbf{p}| + \frac{1}{2}(1 -\eta) V_{\alpha}
+\frac{m_{\alpha}^{2}}{2|\mathbf{p}|}%
+\ldots,\label{largepd2} \\
|\alpha\varsigma\eta \rangle & =|\alpha\rangle 
- \frac{1}{2|\mathbf{p}|}
\sum_{\beta\neq\alpha}|\beta\rangle\frac{1}{V_{(\beta)} -  V_{(\alpha)}}
\langle\beta|\mathbb{M}_{R}\mathbb{M}_{L}|\alpha\rangle +\ldots~~~~(\eta = -1), \label{largepd3}
\end{align}
where $m_{\alpha}^{2}=\langle\alpha|\mathbb{M}_{R}\mathbb{M}_{L}|\alpha
\rangle.$ As noted earlier, the poles of the Majorana
neutrino and Dirac neutrino propagators formally coincide after the energy shift $E \rightarrow
E - \mathbb{V}$ and redefinition $\mathbb{V} \to \mathbb{V}/2$; this is reflected in the dispersion laws
(\ref{major2}) and (\ref{largepd2}). According to Eq. (\ref{largepd2}), a Dirac neutrino with right helicity does not feel the external potential, which physically distinguishes a Dirac neutrino from a Majorana neutrino. 
Moreover, for $\eta = +1$, the term $\mathbb{V}(E -\eta|\mathbf{p}|)$ in
Eq. (\ref{eigenvalueq}) is not dominant, so the validity of the expressions (\ref{largepd1})--(\ref{largepd3}) is restricted to the case $\eta = -1$. 

We omit the intermediate steps and give the final expression for neutrino component of the oscillation amplitudes of Dirac neutrinos
\begin{eqnarray}
\hat{A}(\nu_{\beta L} \leftarrow\nu_{\alpha L}) &=&  - \hat{A}((\nu^c)_{\beta R} \rightarrow (\nu^c)_{\alpha R}) \nonumber \\ &\propto& 
- \sum_{i\eta} e^{-iE_{i\eta}^{D+}(\mathbf{p})\tau}
\langle \beta | \mathbb{F}_{i \varsigma \eta} ( E_{i\eta}^{D+} ,\mathbf{p}) | \alpha \rangle 
\frac{2\hat{p}_{i\eta}^{D+}}{2{E}_{i\eta}^{D+}(\mathbf{p}) - \lambda _{i\eta }^{\prime }({E}_{i\eta}^{D+}(\mathbf{p})) }\Pi_{\varsigma}P_{\eta},  \label{osc5}
\end{eqnarray}
where $\varsigma = + 1$, $\lambda _{i\eta }^{\prime }(E)= {\partial }\lambda
_{i\eta }(E)/{\partial E}$, and $E$ is replaced by $ E_{i\eta}^{D+}$ in $\mathbb{F}_{i \varsigma \eta} ( E,\mathbf{p})$. 
By virtue of Eq.~(\ref{eigenvalueq}) $\lambda _{i\eta }^{\prime }(E) = 
\langle i \varsigma \eta |\mathbb{V}|i \varsigma \eta \rangle $.
Expressions with radicals complicate analysis of the analytical properties of the propagator as a function of energy. Returning to expressions (\ref{simplify bis}) and (\ref{simplify4}), we note that the inversion of matrices in the denominators results in the denominators becoming a polynomial of degree $2n = 6$, which, in the absence of degeneracy, has simple zeros on the real axis of energy. The expression shown in (\ref{osc5}) is the result of integration with respect to the time-like variable $p^0$, while the integration contour is closed in the lower half-plane, where the propagator has no singularities. 

\section{Other neutrino-related processes}
\renewcommand{\theequation}{IV.\arabic{equation}}
\setcounter{equation}{0}

It was demonstrated in the preceding sections that experiments involving neutrino oscillations in a medium and in a vacuum can both be explained solely in terms of the neutrino mass matrix without the need for the mixing matrix. This is demonstrated via Sylvester's theorem, which makes use of Frobenius covariants. Additional neutrino-related processes are covered below. It turns out that the probabilities are also directly expressed using the neutrino mass matrix in all of the scenarios that are considered to be the most typical.

\subsection{$\beta$-decay of tritium}

Tritium is used as a $\beta$ emitter in advanced experiments on the direct measurement of the mass
of an electron antineutrino \cite{Kraus:2005,Aseev:2011,KATRIN:2022}. For the last decade, precise measurements of the electron antineutrino mass have been carried out by the KATRIN collaboration. As a result of the analysis of the edge of the experimental spectrum, a restriction on the neutrino mass $ < 0.8$ eV at CL = 90\% was obtained \cite{KATRIN:2022}.

The edge of the $\beta$ electron energy spectrum is not sensitive to decay channels associated with the electron shell excitation. However, when the $\beta$ electron energy exceeds the binding energy of an electron in a tritium atom, it is necessary to take into account new channels available in energy. The loss of energy for ionization leads to a distortion of the shape of the $\beta$ spectrum. In the $\beta$ decay of tritium, the Coulomb interaction of the knocked-out electron with the nucleus and the channels associated with the population of higher discrete levels by an atomic electron are also important to construct a realistic $\beta$ decay spectrum over the entire range of the energy spectrum of $\beta$ electrons.
Theoretical corrections arising on the atomic and nuclear level are also discussed in ~\cite{Mertens:2015,Tyirn:2019}.

Neglecting electron shell structure effects the $\beta $-decay
probability of tritium equals (see, e.g., \cite{Okun:1981}) 
\begin{equation}
d\Gamma _{\beta }=\frac{G_{\mathrm{F}}^{2}\cos ^{2}\Theta_{\mathrm{C}}}{2\pi ^{3}}|\mathcal{M}_{\mathrm{%
nucl}}|^{2}F(E,Z=2)\sum_{i}
\left\vert \langle i+|\alpha \rangle \right\vert^{2}
p_{\nu _{i}}E_{\nu _{i}}p_{e}E_{e}dE_{e},
\end{equation}%
where $G_{\mathrm{F}}$ denotes the Fermi constant, $\Theta_{\mathrm{C}}$ is the Cabibbo
angle, $\mathcal{M}_{\mathrm{nucl}}$ is the nuclear matrix element, $F(E,Z=2)$ is the
Fermi function, $p_{\nu _{i}}$ and $E_{\nu _{i}}$ are the momentum and
energy of the neutrino with a mass of $m_{i}$, $p_{e}$ and $E_{e}$ are the
momentum and energy of the $\beta $ electron, $Q=E_{\nu _{i}}+E_{e}$ is the
decay energy, $|\alpha \rangle $ is the electron-flavor neutrino state,
and $ \langle i+|\alpha \rangle $ is the mixing matrix.
The kinematic factor is the phase space volume. Using the formula (\ref%
{expl7}), we obtain 
\begin{equation}
d\Gamma _{\beta }=\frac{G_{\mathrm{F}}^{2}\cos ^{2}\Theta_{\mathrm{C}}}{2\pi ^{3}}|\mathcal{M}_{\mathrm{%
nucl}}|^{2}F(E,Z=2)\sum_{i}\langle \alpha | \mathbb{F}_{i +} |\alpha \rangle
p_{\nu _{i}}E_{\nu _{i}}p_{e}E_{e}dE_{e}.
\end{equation}%
The decay probability is expressed in terms of the neutrino mass
matrix in the flavor basis.

\subsection{Electron capture}
The inverse to the $\beta$ decay of an atom is the capture of an orbital electron. The following formula for the total probability is used in the limit of zero width of excited electron shells \cite{Behrens:1982}: 
\begin{equation}
\lambda _{x}=\frac{G_{\mathrm{F}}^{2}\cos ^{2}\Theta_{\mathrm{C}}}{(2\pi )^{2}}n_{x}\mathcal{B}_{x}\beta
_{x}^{2}C_{x}\sum_{i}
\left\vert \langle i+|\alpha \rangle \right\vert^{2}
p_{\nu _{i}}E_{\nu _{i}},  \label{eq:lambda}
\end{equation}%
where $n_{x}$ is the relative occupation number of the $x$ shell, 
$\beta _{x}$ is proportional to the wave 
function of the captured electron inside the parent nucleus. 
The overlap-exchange correction $\mathcal{B}_{x}$ is a square of 
the atomic transition matrix element which takes into account collective electron effects. The phase-space volume is proportional to $p_{\nu_i }E_{\nu_i}$.
The shape function of the transition, $C_{x}$, contains information 
about the nuclear structure and transformation properties of the decay amplitude. 
Other notations as in the previous subsection. 

In terms of the Frobenius covariant (\ref{expl7}),
the decay probability becomes
\begin{equation}
\lambda _{x}=\frac{G_{\mathrm{F}}^{2}\cos ^{2}\Theta_{\mathrm{C}}}{(2\pi )^{2}}n_{x}\mathcal{B}_{x}\beta
_{x}^{2}C_{x}\sum_{i}
\langle \alpha |  \mathbb{F}_{i +} |\alpha \rangle
p_{\nu _{i}}E_{\nu _{i}}.  \label{eqeq:lambda}
\end{equation}%

Electron capture is used or is planned to be used to measure the electron neutrino mass 
\cite{Gastaldo:2014,Faverzani:2016,Gastaldo:2017,Ge:2022a,Ge:2022b}.

\subsection{Neutrinoless double-$\beta$ decay}

Experimental observation of neutrinoless double-$\beta$ decays ($0\nu \beta^- \beta^- $, $0\nu \beta^+ \beta^+ $), 
neutrinoless double electron capture ($0\nu $2EC) or mixed decays $0\nu \beta^-$EC could prove the
non-conservation of the total lepton number and, accordingly, Majorana
nature of neutrino mass. These processes are also of interest for
determining the absolute scale and the hierarchy type of neutrino masses.
In view of the value of processes sensitive to the non-conservation of the total
lepton number, extensive literature is devoted to the physics of $0\nu \beta
\beta $ decay, $0\nu $2EC, and the corresponding models of the nuclear
structure \cite{Suhonen:2007,Simkovic:2011,Vergados:2012,Raduta:2015,Engel:2017,Ejiri:2019,Blaum:2020}.

The $0\nu \beta^- \beta^- $, $0\nu \beta^+ \beta^+ $, $0\nu \beta^+ $EC decays and $0\nu$2EC are caused by the chirality flip of
a massive Majorana neutrino. The amplitudes of the $0\nu \beta^+ \beta^+ $, $0\nu \beta^+ $EC and $0\nu$2EC  processes are proportional to
the propagator projected onto the left-handed state of the emitted and the left-handed
state of the absorbed neutrino. The projection of the propagator can be
written as follows 
\begin{equation}
\langle \alpha |\Pi _{+}\hat{\mathbb{S}}_{F}(p)\Pi _{+}|\alpha \rangle
=\sum_{i}\frac{1}{p^{2}-m_{i}^{2}}\langle \alpha |\mathbb{M}_{L}
\mathbb{F}_{i+} |\alpha \rangle \Pi _{+},
\label{Lprojplus}
\end{equation}%
where $|\alpha \rangle $ is the electron-flavor state of the neutrino in the
initial and final states.\footnote{\footnotesize{\textcolor{black}{In terms of the PMNS matrix (\ref{PMNS}), Eq.~(\ref{Lprojplus}) reads
\begin{equation*} 
\langle \alpha |\Pi _{+}\hat{\mathbb{S}}_{F}(p)\Pi _{+}|\alpha \rangle
=\sum_{ij\beta }\frac{ m_{j}}{p^{2}-m_{i}^{2}} U_{\alpha j}^{\dagger} U_{\beta j}^{\dagger} U_{\beta i} U_{\alpha i}^{\dagger} \Pi _{+}.
\end{equation*}
}}}
We use the condition of completeness of
mass eigenstates $|i\varsigma \rangle $ and take into account that the
transmitted momentum of neutrino has the order of the Fermi momentum at
saturation $p_{\mathrm{F}}=270$ MeV. Since this momentum is much higher than the
neutrino masses $m_{i}\lesssim 1$ eV, one can replace $m_{i}^{2}$ by the
average value of the squares of the neutrino masses, $\langle
m_{i}^{2}\rangle $. By performing the summation, one has   
\begin{equation}
\langle \alpha |\Pi _{+}\hat{\mathbb{S}}_{F}(p)\Pi _{+}|\alpha \rangle
\approx \frac{1}{p^{2}-\langle m_{i}^{2}\rangle }\langle \alpha |\mathbb{M}%
_{L}|\alpha \rangle \Pi _{+}.
\end{equation}

The decay rate 
 is determined by a single element of the mass matrix: 
\begin{equation}
d\Gamma \varpropto \left\vert\langle \alpha |\mathbb{M}_{L}|\alpha \rangle \right \vert ^{2} 
=\left\vert \sum_{i}m_{i}\langle i+|\alpha \rangle^{2}\right\vert ^{2}.  \label{0n2b}
\end{equation}

Contributions to the double-$\beta $ processes of Majorana neutrinos with 
a mass greater than $p_{\mathrm{F}}$ are also discussed \cite{Babic:2017}. For heavy neutrinos, it is necessary to
perform summation in Eq.~(\ref{Lprojplus}) with the exact Frobenius covariants (\ref{expl7}). 

Equation (\ref{Lprojplus}) shows that the $0\nu \beta \beta $,  $0\nu \beta^-$EC and  $0\nu$2EC probabilities are expressed in terms of the mass matrix alone.

\subsection{Quasi-elastic neutrino scattering}

As the next example, consider a neutrino scattered on a lepton: $\nu_i + \ell_i^- \to \nu_j + \ell_f^-$. According to the scattering theory, there is an asymptotic state 
$| i + \rangle \otimes | \ell_i^- \rangle $ 
of a neutrino and a lepton with masses $m_i$ and $\mu_i$ at infinity $t \to - \infty$, and an asymptotic state $| j + \rangle \otimes | \ell_f^- \rangle $ 
of a neutrino and a lepton with masses $m_j$ and $\mu_f$  
at infinity $t \to + \infty$. Since the interaction is formulated in electroweak basis, 
the amplitude $\mathcal{A}_{\alpha \beta}$ is calculated first for flavor neutrinos, then converted to the basis of the diagonal in mass asymptotic states:
\begin{equation}
\mathcal{A}_{ji} = \langle j +|\alpha \rangle \mathcal{A}_{\alpha \beta} \langle \beta |i + \rangle.
\end{equation}
The summation is implied over the repeated indices. The cross section is represented as
\begin{equation}
d\sigma =\frac{1}{4\sqrt{(p_{i}p_{\ell^-_i})^{2}-m_{i}^{2}\mu_i^{2}}}
\left\vert \mathcal{A}_{ji}\right\vert ^{2}
\frac{1}{(2\pi )^{2}}\Phi _{2}(\sqrt{s},m_{j},\mu_f),
\end{equation}
where $p_{\ell^-_i}$ is the initial four-momentum of lepton, $s = (p_i + p_{\ell^-_i})^2$, and $\Phi _{2}(\sqrt{s},m_{j},\mu_f)$ is the relativistic phase space of the final particles. 
The identity $|\mathcal{A}_{ji}|^2 = \mathrm{Tr}\left[ \mathcal{A} |i + \rangle \langle i + | \mathcal{A}^{\dag }|j +\rangle \langle j+|\right]$ allows the cross section for the non-polarized particles to be written in terms of the Frobenius covariants (\ref{expl7}):
\begin{eqnarray}
d\sigma  
=\frac{1}{4\sqrt{(p_{i}p_{\ell^-_i})^{2}-m_{i}^{2}\mu_i^{2}}}
\mathrm{Tr}\left[\mathcal{A}  \mathbb{F}_{i +}
\mathcal{A}^{\dag } \mathbb{F}_{j +} \right] \frac{1}{(2\pi )^{2}}\Phi _{2}(\sqrt{s},m_{j},\mu_f).
\end{eqnarray}%
The trace can be calculated in flavor basis. Explicit use of a basis diagonal in neutrino masses is not mandatory. 
The corresponding expressions under the trace sign contain only the neutrino mass matrix. These conclusions are valid for both Majorana neutrinos and Dirac neutrinos.

\vspace{2mm}
After completing this paper, our attention was drawn to the papers \cite{Bellandi:1997,Aquino:1999,Fogli:1999,Denton:2020,Denton:2022}, 
where the problem of constructing the mixing matrix in terms of the mass matrix is discussed, 
including the use of analogues of Frobenius covariants in a less general framework.
Equation (28) of the paper \cite{Fogli:1999}
is reproduced from Eqs.~(\ref{eigenstate1}) and (\ref{eigenvalueq}) with the help of perturbation theory 
in the limit of small mean-field potential $|\mathbf{p}||V_{\alpha}| \ll m_i^2$ and $|\mathbf{p}| \gg m_i$. 
In Eqs.~(\ref{major3}) and (\ref{largepd3}) we give similar expressions in the limit of
$|\mathbf{p}||V_{\alpha}| \gg m_i^2$ and $|\mathbf{p}| \gg m_i$, which is interesting for physics of newly born 
neutron stars. 
In Ref.~\cite{ Denton:2022}, an interesting theorem is discussed on the relashionship of the module 
of the mixing matrix elements with the eigenvalues of the mass matrix and the eigenvalues of the diagonal minors 
of the mass matrix. 
The Frobenius covariants, apparently, successfully solve the problem of constructing the mixing matrix,
including determining the phases of matrix elements. In the simplest case of neutrinos in vacuum, 
Eqs.~(\ref{expl7}) and (\ref{PMNS}) give, in particular,
\begin{equation}
U_{\alpha i} U_{\beta i}^* = \langle \alpha | \mathbb{F}_{i+}| \beta \rangle.
\end{equation}

The main message of our paper,  on the other hand, is that in neutrino physics, all physical characteristics of the processes are expressed through the neutrino mass matrix without reference to the mixing matrix or assumptions about the hierarchy of neutrino masses. The proposed procedure works for the Majorana mass 
term, a complex symmetric matrix (\ref{massmatrix}) specified by nine parameters for three neutrino flavors.
Of course, an observation of the $0\nu\beta\beta$-decay or an assumption about the value of $M^M_{ee}$ is necessary. 

\section{Conclusion}
\renewcommand{\theequation}{V.\arabic{equation}}
\setcounter{equation}{0}

In light of the lack of an explanation for the observed violation of $CP$ symmetry, the abundance of baryons in the universe, the presence of dark matter, etc., the limitations of SM are obvious. The characteristics of neutrinos, such as their nature (Majorana or Dirac), and the mass matrix with a large number of parameters, can point the right path when looking for SM generalizations. This paper demonstrates that the amplitudes of neutrino-related processes can be constructed by methods, which do not require a mixing matrix and instead allow explicit representations to be built purely in terms of the neutrino mass matrix. 

Sections II and III analyze in detail the involved processes of neutrino oscillations in a vacuum and in a medium. Section IV demonstrates that the probabilities of the most typical processes described by diagrams with external neutrino lines are also expressed in terms of only the neutrino mass matrix without reference to the mixing matrix. We discuss quasi-elastic neutrino scattering, and single- and double-beta decays. Although the scattering theory is formulated in terms of asymptotic states, the transformation of scattering amplitudes into the basis of massive diagonal states is not mandatory. 

Compared to methods using a mixing matrix, the neutrino mass matrix can be sensitive to alternative experimental data sets.
Indirect determination of the mass matrix through the mixing matrix involves loss of information, which leads typically to an increase in statistical errors. 
The components of the neutrino mass matrix can apparently be determined with reduced errors if the measurement data are directly compared with the mass matrix. \textcolor{black}{Using such a fitting, it is potentially possible to gain access to the structure of the mass matrix - Majorana or Dirac.}

\begin{acknowledgments}
The authors are grateful to Prof.~E.~Lisi for reading the manuscript and valuable comments.
\end{acknowledgments}

\end{document}